%% file: GenerativeControl.tex
\documentclass[acmtog,authorversion]{acmart}

\input{Sections/_template_args.tex}
\input{Sections/_preamble.tex}

\begin{document}
\title{{ControlVAE: Model-Based Learning of Generative Controllers for Physics-Based Characters}}

\input{Sections/0_authors.tex}

\input{Sections/0_abstract.tex}

%
%

 
%
%
\input{Sections/0_teaser.tex}

\maketitle

\input{Sections/1_Introduction.tex}

\input{Sections/2_Related_Work.tex}
\input{Sections/4_ControlVAE_LB}

\input{Sections/5_High_Level_Control.tex}
\input{Sections/7_Results}

\input{Sections/8_Discussion}

\input{Sections/x_ack.tex}

\input{Sections/x_appendix}
\bibliographystyle{ACM-Reference-Format}
\bibliography{GenerativeControl}


\end{document}

%% file: Sections/_template_args.tex
\AtBeginDocument{%
  \providecommand\BibTeX{{%
    \normalfont B\kern-0.5em{\scshape i\kern-0.25em b}\kern-0.8em\TeX}}}

\setcopyright{acmcopyright}
\acmJournal{TOG}
\acmYear{2022}
\acmVolume{41}
\acmNumber{6}
\acmArticle{183}
\acmMonth{12}
\acmDOI{10.1145/3550454.3555434}

\acmSubmissionID{144}



%% file: Sections/_preamble.tex
\usepackage{bm}

\usepackage{graphicx}
\usepackage[labelformat=simple]{subcaption}
\usepackage{amsmath}
\usepackage{float}
\usepackage{soul}
\usepackage{mathrsfs}

\newcommand{\vect}[1]{\bm{#1}}

\newcommand{\eqword}[1]{{\text{#1}}}

\usepackage{booktabs} 

\usepackage[ruled]{algorithm2e} 

\SetAlFnt{\small}
\SetAlCapFnt{\small}
\SetAlCapNameFnt{\small}
\SetAlCapHSkip{0pt}

\usepackage{cases}

\usepackage{xspace}
\newcommand{\ie}{\textsl{i.e.}~}
\newcommand{\eg}{\textsl{e.g.}~}

\newcommand{\keep}[1]{}
\newcommand{\old}[1]{}

\DeclareMathOperator*{\argmin}{arg\,min}

\DeclareMathOperator*{\diag}{diag}
\newcommand{\loss}{\ensuremath{\mathcal{L}}}
\newcommand{\E}{\ensuremath{\mathbb{E}}}
\newcommand{\KL}[2]{\ensuremath{\mathcal{D}_{\text{KL}}\left(#1 \Vert #2\right)}}

\newcommand{\dataset}{\ensuremath{\mathcal{D}}}
\newcommand{\simBuffer}{\ensuremath{\mathcal{B}}}
\newcommand{\cvae}{ControlVAE}
\newcommand{\svae}{VAE with non-conditional prior}
\newcommand{\svaeShort}{VAE-NC}

\newcommand{\fig}{Figure{}~}
\newcommand{\eqn}{Equation{}~}
\newcommand{\Sec}{Section{}~}
\newcommand{\Tab}{Table{}~}

\newcommand{\stt}{\ensuremath{\vect{s}}}
\newcommand{\act}{\ensuremath{\vect{a}}}
\newcommand{\policy}{\ensuremath{{\pi}}}
\newcommand{\latent}{\ensuremath{\vect{z}}}
\newcommand{\Latent}{\ensuremath{\mathcal{Z}}}
\newcommand{\task}{\ensuremath{\vect{g}}}

%% file: Sections/0_authors.tex
\author{Heyuan Yao}
\email{heyuanyao@pku.edu.cn}
\orcid{0000-0002-6168-6777}
\affiliation{%
  \institution{SCS \& KLMP (MOE), Peking University}
  \streetaddress{No.5 Yiheyuan Road, Haidian District}
  \city{Beijing}
  \state{Beijing}
  \country{China}
  \postcode{100871}
}

\author{Zhenhua Song}
\email{songzhenhua@stu.pku.edu.cn}
\orcid{0000-0001-5139-209X}
\affiliation{%
  \institution{SCS \& KLMP (MOE), Peking University}
  \streetaddress{No.5 Yiheyuan Road, Haidian District}
  \city{Beijing}
  \state{Beijing}
  \country{China}
  \postcode{100871}
}

\author{Baoquan Chen}
\email{baoquan@pku.edu.cn}
\orcid{0000-0003-4702-036X}  
\affiliation{%
  \institution{SIST \& KLMP (MOE), Peking University}
  \streetaddress{No.5 Yiheyuan Road, Haidian District}
  \city{Beijing}
  \state{Beijing}
  \country{China}
  \postcode{100871}
}

\author{Libin Liu}
\authornote{corresponding author}
\email{libin.liu@pku.edu.cn}
\orcid{0000-0003-2280-6817}
\affiliation{%
  \institution{SIST \& KLMP (MOE), Peking University}
  \streetaddress{No.5 Yiheyuan Road, Haidian District}
  \city{Beijing}
  \state{Beijing}
  \country{China}
  \postcode{100871}
}
\renewcommand{\shortauthors}{Yao, Song, Chen, and Liu}

%% file: Sections/0_abstract.tex
\begin{abstract}
    In this paper, we introduce ControlVAE, a novel model-based framework for learning generative motion control policies based on variational autoencoders (VAE). Our framework can learn a rich and flexible latent representation of skills and a skill-conditioned generative control policy from a diverse set of unorganized motion sequences, which enables the generation of realistic human behaviors by sampling in the latent space and allows high-level control policies to reuse the learned skills to accomplish a variety of downstream tasks. In the training of ControlVAE, we employ a learnable world model to realize direct supervision of the latent space and the control policy. This world model effectively captures the unknown dynamics of the simulation system, enabling efficient model-based learning of high-level downstream tasks. We also learn a state-conditional prior distribution in the VAE-based generative control policy, which generates a skill embedding that outperforms the non-conditional priors in downstream tasks. We demonstrate the effectiveness of ControlVAE using a diverse set of tasks, which allows realistic and interactive control of the simulated characters.
\end{abstract}

\keywords{physics-based character animation, motion control, deep reinforcement learning, generative model, VAE}

%% file: Sections/0_teaser.tex
\begin{teaserfigure}
    \centering
    \includegraphics[width=7.0in]{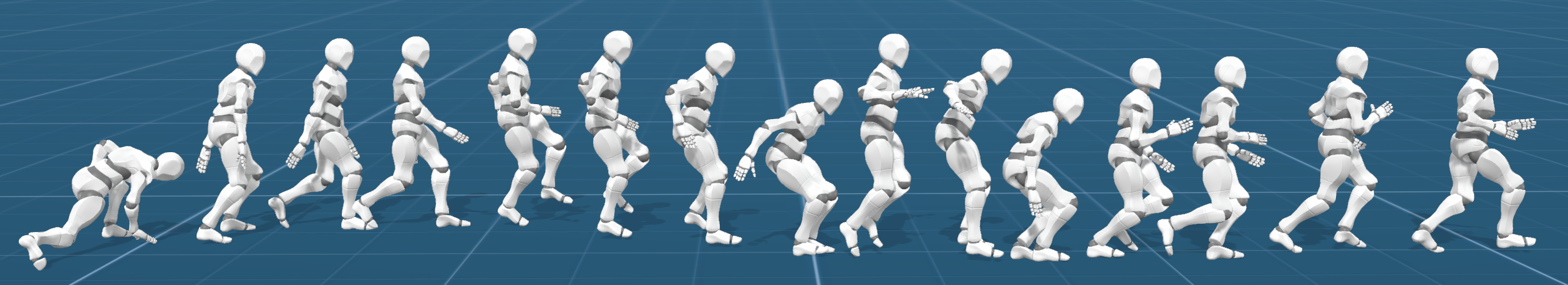}
    \caption{A simulated character gets up, walks, hops, jumps, and runs in our system. We demonstrate a model-based framework for learning a generative motion control policy based on variational autoencoders (VAE), which allows a simulated character to learn a diverse set of skills and use them to accomplish various downstream tasks.}
  \end{teaserfigure}

%% file: Sections/1_Introduction.tex
\section{Introduction}
Learning physics-based controllers to realize complex human behaviors has been a longstanding challenge for character animation. Recent research has partially addressed this problem by imitating motion capture data of real human performance using deep reinforcement learning. However, there is a common practice in these approaches where control policies for different tasks are often trained from scratch. Although an extensive range of motions, from basic locomotion to dynamic stunts, have been successfully learned by various agents, how to effectively reuse these learned motion skills to accomplish new tasks remains a challenging problem.


Recent research in kinematic motion synthesis has demonstrated successful applications of generative models, such as variational autoencoders (VAE)~\cite{lingCharacterControllersUsing2020a} and normalizing flows~\cite{HenterMoGlow2020}, in learning a rich and versatile latent space to encode a large variety of skills that multiple downstream tasks can reuse. However, these approaches are not directly applicable to learning generative physics-based controllers because their learning objectives are usually defined in the motion domain. The gradients of these objectives typically cannot be back-propagated over the barrier of the simulation, which is often considered as a black box and not differentiable. Even with a differentiable dynamics engine, the highly non-linear nature of the articulated rigid-body system and the contact dynamics often cause the training to stop at poor local optima \cite{werlingDiffDart2021,Hamalainen2020_Visualizing}.
A possible way to sidestep this issue is to convert the learning objective of the generative model into a special reward term of reinforcement learning~\cite{pengAMPAdversarialMotion2021a,PengASELargescaleReusableAdversarial2022,HoGailNips2016}. However, this approach will require training two coupled indirectly supervised systems, where the training process and the training objective often need to be carefully designed to achieve stable results.


Model-based approaches, or, more specifically, those learning world models~\cite{WorldModelHa2018}, incorporate approximate models to predict the future states of the system given its previous states and the actions taken. The differentiable world model bridges the gap between the simulation and control policy, allowing the objectives defined in the motion domain to supervise the policy directly, thus achieving efficient and stable training \cite{Deisenroth2011_PILCO,Janner2021_MBPO}. Recently, model-based reinforcement learning has shown promising results in learning to track complex human motions \cite{fussellSuperTrackMotionTracking2021}. The success of these studies suggests a possibility that a more flexible control model, such as a generative model like VAE, can be learned with the help of a learnable world model.


In this paper, we introduce \cvae{}, a novel model-based framework for learning generative motion control policies based on variational autoencoders (VAE). Our framework can learn a rich and flexible latent representation of skills and a skill-conditioned generative control policy from a diverse set of unorganized motion sequences, which enables the generation of realistic human behaviors by sampling in the latent space and allows high-level control policies to reuse the learned skills to accomplish a variety of tasks.

VAEs are commonly trained against the standard normal distribution \cite{Kingma2014_VAE,lingCharacterControllersUsing2020a}. However, our preliminary experiments show that such a non-conditional prior is not efficient in disentangling the representations of different skills. We thus model our VAE-based generative policy with a prior distribution conditional on the simulated character state. The latent skill embeddings generated from this conditional prior distribution achieve higher performance in downstream tasks than those learned using the non-conditional priors.

We employ a learnable world model to approximate the unknown dynamics of the simulation system, which is trained using online samples. We find this world model not only provides direct supervision for learning the latent space and the control policy, but further allows efficient model-based optimization of control policies of downstream tasks with model-based learning. 



To evaluate our method, we train \cvae{} on a diverse set of locomotion skills and test its performance on several challenging downstream tasks. We further conduct studies to validate our design decisions both qualitatively and quantitatively.



    


%% file: Sections/2_Related_Work.tex
\section{Related Work}

\subsection{Physics-based Motion Controllers}
Research on developing physics-based control strategies to realize realistic and interactive motions has a long history in computer animation. The seminal works can be dated back to the 1990s, when locomotion control was realized based on careful motion analysis and hand-crafted controllers \cite{Hodgins1995_Animating}. Robust locomotion controllers are later developed using abstract models \cite{corosGeneralizedBipedWalking2010,yinSIMBICONSimpleBiped2007,Lee2010_Datadriven}, optimal control \cite{Muico2011_Composite}, model predictive control \cite{Mordatch2010_Robust,Hamalainen2015_Online}, policy optimization \cite{tanLearningBicycleStunts2014}, and reinforcement learning \cite{yuLearningSymmetricLowenergy2018,YinJumping2021,XieALLSTEPS2020}. These approaches typically require sufficient prior knowledge and hand-tuned parameters or reward functions, hence can be hard to apply to complex motions and scenarios.
Such difficulties motivate the so-called data-driven methods that generate natural motion by imitating human performance. A fundamental task of these approaches is to track a reference motion by learning feedback policies~\cite{liuTerrainRunnerControl2012,liuGuidedLearningControl2016,Lee2010_Datadriven}. The development of deep reinforcement learning techniques further enables robust tracking of agile human motions \cite{pengDeepMimicExampleguidedDeep2018a} and to generalize to various body shapes \cite{WonBodyShapeVariation2021} and environments \cite{XieALLSTEPS2020}. 

Based on the success of individual controllers, recent research starts to focus on creating multi-skilled characters.
Training a robust tracking policy to track the results of a pre-trained kinematic controller allows the combined control strategy to imitate different target motions according to the task or user input \cite{bergaminDReConDatadrivenResponsive2019,ParkPredictandSimulate2019,wonScalableApproachControl2020}. However, the performance of such a combined strategy is limited to the capability of the kinematic controller, and it can be computationally expensive to evaluate both the networks of the tracking policy and kinematic controller at runtime.  
Individual physics-based control policies can be organized into a graph-like structure \cite{liuGuidedLearningControl2016,pengDeepMimicExampleguidedDeep2018a}, which can be further broken down into short snippets and managed by a high-level scheduler \cite{liuLearningScheduleControl2017}. However, the system needs to maintain all the sub-controllers at runtime for downstream tasks. 
More recent studies \cite{merelNeuralProbabilisticMotor2018,merelCatchCarryReusable2020,luoCARLControllableAgent2020,pengAMPAdversarialMotion2021a,PengASELargescaleReusableAdversarial2022,Won2022ConditionalVAE} develop various generative models to incorporate a diverse set of motions. The results are compact latent representations of skills that allow high-level policies to reuse multiple skills to accomplish downstream tasks. However, learning efficient representations of skills is a nontrivial problem.
We develop several novel components in this work to ensure the performance of the learned skill embeddings in the downstream tasks, which are verified in a series of experiments.





\subsection{Generative Models in Motion Control}
Generative models have been extensively studied in kinematic motion synthesis. Early research learns statistical representation of motion based on the Gaussian Mixture Model \cite{MotionGraphPPChaiJinXiang2012} and Gaussian Process \cite{Wang2008Gaussian,LevineGaussianContinuousCharacterControl2012}. In the era of deep learning, the VAEs \cite{Kingma2014_VAE} have been realized using the mixture-of-expert network \cite{lingCharacterControllersUsing2020a} and transformers \cite{PetrovichACTORTransformerVAE2021} to encode motions into latent spaces. The MoGlow framework proposed by \citet{HenterMoGlow2020} draws support from the normalizing flows \cite{Kingma2018_GLOW} for probabilistic motion generation. \citet{Li2022_GANimator} train a GAN-based model to synthesize convincing long motion sequences from a short example. 
A generative model usually learns a compact latent space that encodes a variety of motions. Sampling in the latent space leads to natural kinematic motions, which allows a task-oriented high-level controller to reuse the motions in downstream tasks using either optimization \cite{MotionGraphPPChaiJinXiang2012,Holden2016MotionManifold} or learned policies \cite{LevineGaussianContinuousCharacterControl2012,lingCharacterControllersUsing2020a}.

Physics-based controllers can also be formulated as generative models, where the samples in the latent space will be decoded into specific actions for the simulated character.
Merel et al.~\shortcite{merelNeuralProbabilisticMotor2018} learn an autoencoder that distills a large group of expert policies into a latent space, with which a high-level controller can be learned to accomplish difficult tasks like catching and carrying an object \cite{merelCatchCarryReusable2020}. 
Similarly, \citet{Won2022ConditionalVAE} employ a conditional VAE and perform behavior cloning on expert trajectories.
\citet{luoCARLControllableAgent2020} treat the composing weights of the motor primitives learned by the multiplicative compositional policies (MCP) \cite{pengMCPLearningComposable2019} as the representation of skills and achieves interactive locomotion control of a simulated quadruped.
Inspired by the generative adversarial imitation learning (GAIL) \cite{HoGailNips2016}, Peng et al.~\shortcite{pengAMPAdversarialMotion2021a} include the adversarial motion prior into the reward function of reinforcement learning, encouraging the simulated character to finish a task using natural-looking actions. They later extend this framework to learn adversarial skill embeddings for a large range of motions \cite{PengASELargescaleReusableAdversarial2022}.
Our \cvae{} is based on the conditional VAE. Unlike previous works using a standard normal distribution as the prior distribution of VAE \cite{lingCharacterControllersUsing2020a,Won2022ConditionalVAE}, we employ a state-conditional prior that creates better latent embeddings. We also employ a model-based learning algorithm to learn our \cvae{} model.

\subsection{Model-based Learning}
Model-based learning approaches realize control using the underlying model of a dynamic system. The models can be either accurate or approximated using learnable functions. Accurate models often come with differentiable simulators \cite{Todorov2012_MuJoCo,werlingDiffDart2021,mora2021pods}. They can be used in either offline or online optimization to construct motion controllers \cite{Mordatch2012_Discovery,mora2021pods,Mordatch2010_Robust,Muico2011_Composite,Macchietto2009_Momentum,Hong2019_Physicsbased,Eom2019_Model}. Approximate models, or the \emph{World Model}s \cite{WorldModelHa2018}, are typically formulated as Gaussian Process \cite{Deisenroth2011_PILCO} or neural networks \cite{AnushaNeuralNetworkDynamics2018,Janner2021_MBPO}. They provide  differentiable transition functions that allow gradients of learning objectives to pass through the barrier of the simulation \cite{schmidhuber1990making,RecurrentEnvSimu2017,HeessSVGControlNIPS2015}, thus enabling policy optimization to be solved efficiently using gradient-based techniques \cite{Deisenroth2011_PILCO,Janner2021_MBPO,HeessSVGControlNIPS2015,AnushaNeuralNetworkDynamics2018}.

The application of the reinforcement learning algorithms with learnable world models is rather sparse in physics-based character animation. While early studies in this category had been done in the 1990s \cite{GrzeszczukNeuroAnimator1998}, SuperTrack~\cite{fussellSuperTrackMotionTracking2021} is among the first frameworks that achieve successful tracking of a large diversity of skills. Our work is inspired by SuperTrack~\cite{fussellSuperTrackMotionTracking2021}. We also learn a world model along with the control policy to achieve efficient training. However, unlike SuperTrack~\cite{fussellSuperTrackMotionTracking2021}, which learns individual tracking controllers, our system learns a VAE-based generative control policy and reusable skill embeddings, which enable multiple downstream tasks without training from scratch each time.

A concurrent study done by \citet{Won2022ConditionalVAE} shares a similar goal to our work. They also develop a VAE-based control policy and learn a world model to facilitate training. Our method differs from their approach in three ways: (a) \citet{Won2022ConditionalVAE} employ behavior cloning to train the control policy, which requires the demonstration of many pre-trained expert policies. Instead, our framework allows direct learning of the generative control policies from raw motion clips; (b) our method trains the world model along with the control policy, ensuring they are compatible. We believe this is critical for successful model-based learning of the downstream tasks. The same results are not demonstrated by \citet{Won2022ConditionalVAE}. The world model is learned offline in their system; (c) we employ a state-conditional prior distribution in our VAE-based model, which outperforms the non-conditional prior in accomplishing downstream tasks. In contrast, \citet{Won2022ConditionalVAE} use the standard state-independent prior.

%% file: Sections/4_ControlVAE_LB.tex
\section{Control VAE}

\begin{figure*}
    \centering
    \includegraphics[width=0.8\linewidth]{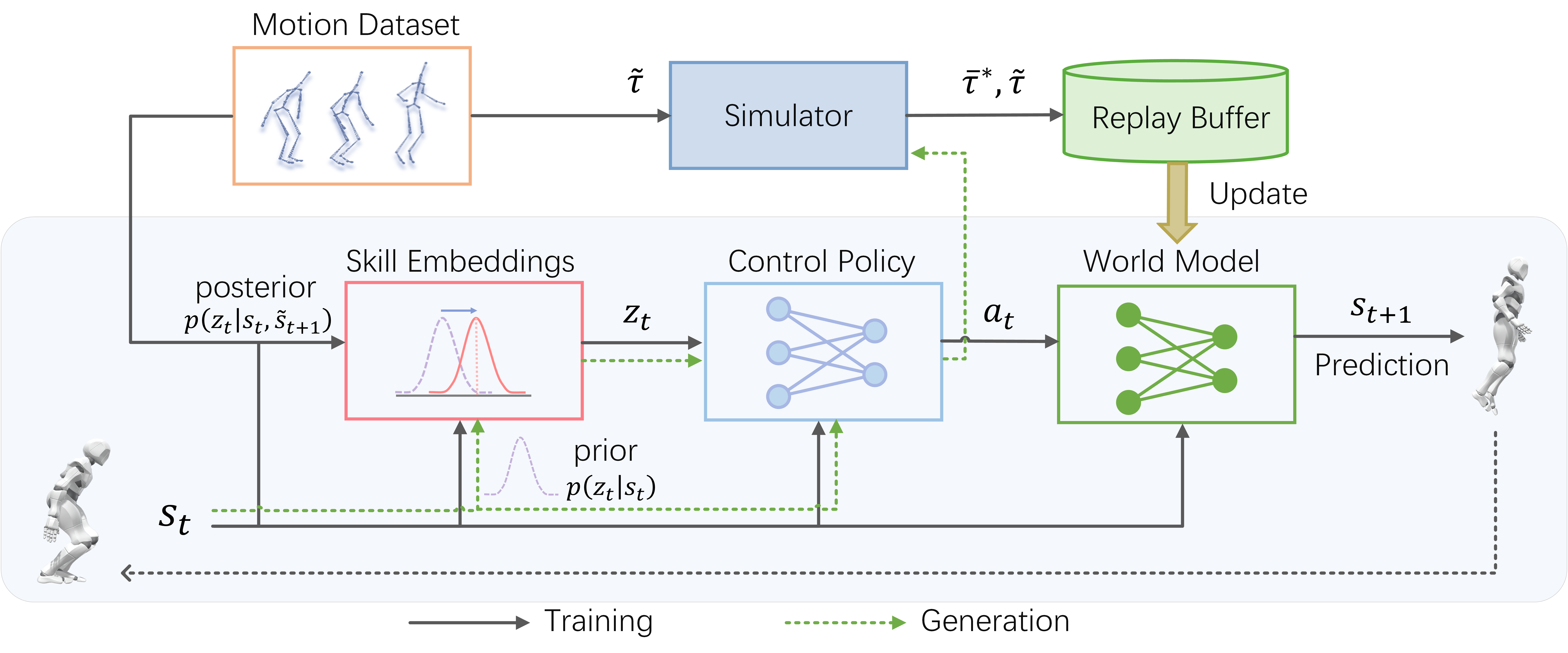}
    \caption{Overview of our ControlVAE System. }
    \label{Fig:Pipeline}
\end{figure*}

Figure \ref{Fig:Pipeline} illustrates the overall structure of \cvae{}.
In \cvae{}, a motion control policy is formulated as a conditional distribution $\policy(\act|\stt,\latent)$. It computes an action $\act$ to control the simulated character according to its current state $\stt$ and a latent variable $\latent\in\Latent$ sampled from a prior distribution $\latent\sim{}p(\latent)$. We train this control policy to allow each latent variable $\latent$ to encode a specific skill, which enables a high-level policy to operate in the latent space $\Latent$ to accomplish downstream tasks. 

More specifically, at each time step $t$, the character observes its state~$\stt_t$ and computes an action $\act_t \sim \policy(\act_t|\stt_t,\latent_t)$, where the latent variable $\latent_t$ {is provided by a high-level policy}. The character then executes $\act_t$ in the simulation and moves to a new state $\stt_{t+1}$ with the probability of transition given by $p(\stt_{t+1}|\stt_t,\act_t)$. This process is then repeated, resulting in a simulated trajectory $\tau^*=\{\stt_0, \act_0, \stt_1, \act_1, \dots, \stt_{T-1}, \act_{T-1}, \stt_T\}$ in which the character keeps moving and performs the skills defined by the sequence of latent codes $\{\latent_{0:T-1}\}$. Here we assume that $\{\latent_{0:T-1}\}$ are generated independently of each other.

What we are interested in is the character's motion in this simulated trajectory, represented compactly as $\tau=\{\stt_{0:T}\}$. To produce realistic skills, we need to train the control policy $\policy(\act_t|\stt_t, \latent_t)$ so that the distribution of the generated motions $p(\tau)$ matches the distribution of a motion dataset $\dataset=\{\tilde{\tau}_i\}$. However, the computation of $p(\tau)$ relies on a very complex likelihood $p(\stt_{0:T}|\latent_{0:T-1})$, 
\begin{align}
    p({\tau}) 
    &= \int_{\latent_{0:T-1}} p(\latent_{0:T-1}) p(\stt_{0:T}|\latent_{0:T-1}),
    \label{eqn:p_tau}
\end{align}
which makes the computation intractable. The variational autoencoder (VAE)~\cite{Kingma2014_VAE} thus considers its evidence lower bound (ELBO). Considering that $\stt_{t+1}$ and $\act_t$ depend only on $\stt_t$ and $\latent_t$,  also $\{\latent_t\}$ are independent to each other, the ELBO of \eqn\eqref{eqn:p_tau} can be written as
\begin{align}    
    \log p({\tau}) - \log p(\stt_0) 
    &\ge \sum_{t=0}^{T-1} \E_{q(\latent_t|\stt_{t},\stt_{t+1})}\left[\log p(\stt_{t+1}|\stt_t,\latent_t)\right] 
    \nonumber\\
    & - \sum_{t=0}^{T-1} \KL{q(\latent_t|\stt_{t},\stt_{t+1})}{p(\latent_t)},
    \label{eqn:ELBO} 
\end{align}
where $q(\latent_t|\stt_{t},\stt_{t+1})$ is the so-called variational distribution, which approximates the true posterior distribution of the skill variable $p(\latent_t|\tau)$ and is assumed to depend on only the state transition, denoted as $(\stt_{t}, \stt_{t+1})$, $p(\stt_{t+1}|\stt_t,z_t)$ represents the probability of this state transition conditional on~$\latent_t$, and $p(\stt_0)$ is the distribution of the start states.

The ELBO essentially defines an autoencoder, where the variational distribution $q(\latent_t|\stt_{t},\stt_{t+1})$ describes the encoding process that a skill representation $\latent_t$ is extracted from a pair of consecutive states $(\stt_t, \stt_{t+1})$, and $p(\stt_{t+1}|\stt_t,z_t)$ characterizes the decoding process where $\stt_{t+1}$ is reconstructed from the skill embedding $\latent_t$. 
The KL-divergence in \eqn\eqref{eqn:ELBO} can be considered as a regularization that encourages the approximate posterior to be close to the prior $p(\latent)$, so that each sample drawn from $p(\latent)$ can be decoded into a realistic skill. 

\subsection{Generation}

A common choice of the prior $p(\latent)$ is the standard multivariate normal distribution $\mathcal{N}(\vect{0},\vect{I})$ \cite{lingCharacterControllersUsing2020a,Kingma2014_VAE}. However, in our preliminary experiments with this state-independent prior, the character often keeps changing skills quickly, leading to jerky movements and occasional falls. Downstream control policies also perform less efficiently on this prior. For example, the character can encounter difficulties in maintaining a constant moving direction in a direction control task. We presume that this state-independent prior distribution may not provide enough information for the regularization. The encoder thus computes an inconsistent skill distribution, where the same skill is encoded into different parts of the latent space at different states.

To deal with this problem, we leverage a conditional distribution $p(\latent_t|\stt_t)$ as the prior 
and formulate it as a Gaussian distribution with diagonal covariance
\begin{align}
    p(\latent_t|\stt_t) \sim \mathcal{N}\left(\vect{\mu}_p(\stt_t;{\theta}_p), \sigma_p^2\vect{I}\right),
    \label{eqn:prior_dist}
\end{align}
where the mean $\vect{\mu}_p$ is a neural network parameterized with $\theta_p$ and the standard deviation $\sigma$ is a hyperparameter. 

We then generate a motion trajectory according to a sequence of skill codes $\{\latent_t\}$ by simulating the character with the actions computed by the policy $\policy(\act_t|\stt_t, \latent_t)$. 
We model the policy as a Gaussian distribution
\begin{align}
    \policy(\act_t|\stt_t,\latent_t) \sim \mathcal{N}(\vect{\mu}_{\pi},\Sigma_{\pi})
\end{align}
with a diagonal covariance matrix $\vect{\Sigma}_{\pi}$ and the mean $\vect{\mu}_{\pi}$ as a neural network $\vect{\mu}_{\pi}(\stt_t, \latent_t;\theta_{\pi})$ parameterized by $\theta_{\pi}$. The generation process is then characterized by the log-likelihood in \eqn\eqref{eqn:ELBO}
\begin{align}
    \log p(\stt_{t+1}|\stt_t, \latent_t)&=\log\int_{\act_t} p(\stt_{t+1}|\stt_t, \act_t) \policy(\act_t | \stt_t, \latent_t) \nonumber\\
    &\ge \E_{\policy(\act_t | \stt_t, \latent_t)} \left[\log p(\stt_{t+1}|\stt_t, \act_t)\right].
    \label{eqn:log_likelihood_transition} 
\end{align}
However, the true transition probability distribution $p(\stt_{t+1}|\stt_t, \act_t)$ 
is not known since we consider the simulation process as a black box. We opt to approximate it using a world model $\omega(\stt_{t+1}|\stt_t, \act_t)$, which is another Gaussian distribution
\begin{align}
    \omega(\stt_{t+1}|\stt_t, \act_t) \sim \mathcal{N}(\vect{\mu}_w, \vect{\Sigma}_w)
    \label{eqn:world_model}
\end{align}
with a diagonal covariance matrix $\vect{\Sigma}_w$ and the mean $\vect{\mu}_w$ computed as a neural network $\vect{\mu}_w(\stt_t, \act_t;\theta_w)$ parameterized by $\theta_w$.


\subsection{Inference}
During the training, we need to infer the approximate posterior distribution of the skill variable $q(\latent_t|\stt_{t}, \stt_{t+1})$ according to each pair of states $(\stt_t, \stt_{t+1})$ in a state transition. Again, we let $q(\latent_t|\stt_{t}, \stt_{t+1})$ be a Gaussian distribution
\begin{align}
    q(\latent_t|\stt_{t}, \stt_{t+1}) \sim \mathcal{N}\left(\hat{\vect{\mu}}_q, \sigma_q^2\vect{I}\right)
    \label{eqn:post_dist}
\end{align}
with a diagonal covariance. 
Considering that $q(\latent_t|\stt_{t}, \stt_{t+1})$ needs to stay close to the prior $p(\latent_t|\stt_t)$, we let $\sigma_q = \sigma_p$. Rather than directly learning the mean~$\hat{\vect{\mu}}_q$, we learn the residual between $\hat{\vect{\mu}}_q$ and the mean of the prior, $\vect{\mu}_p$. Specifically, we compute
\begin{align}
    \hat{\vect{\mu}}_q=\vect{\mu}_p + \vect{\mu}_q ,
\end{align}
where $\vect{\mu}_q=\vect{\mu}_q(\stt_t, \stt_{t+1};\theta_q)$ is a residual network parameterized by $\theta_q$.
This structure is inspired by the SRNN~\cite{fraccaroSequentialNeuralModels2016}. It makes the KL-divergence in \eqn\eqref{eqn:ELBO} independent of the prior $p(\latent_t|\stt_t)$ and allows an efficient computation of
\begin{equation}
    \KL{q(\latent_t|\stt_{t}, \stt_{t+1})}{p(\latent_t|\stt_t)} = \frac{\vect{\mu}_q^T\vect{\mu}_q}{2{\sigma}_p^2}.
    \label{eqn:KL}
\end{equation}
The training process can then be viewed as learning how to correct the prior distribution using the information from the next state $\stt_{t+1}$. 

\subsection{Training}
\subsubsection{Control VAE}
\label{sec:training_control_VAE}
The training of \cvae{} is then formulated as a maximum likelihood estimation (MLE) problem over the motion dataset $\dataset=\{\tilde{\tau}_i\}$. We solve this problem by maximizing the ELBO in \eqn\eqref{eqn:ELBO}, 
where the control policy $\policy(\act_t|\stt_t,\latent_t)$, the conditional prior $p(\latent_t|\stt_t)$, and the approximate posterior $q(\latent_t|\stt_t, \stt_{t+1})$ are trained together. The world model $\omega(\stt_{t+1}|\stt_t, \act_t)$ is learned separately, as will be described later.

Instead of evaluating \eqn\eqref{eqn:ELBO} in a time step-wise manner, we compute it over synthetic trajectories. Specifically, we pick a random start state $\tilde{\stt}_0$ and generate a synthetic trajectory that tracks the corresponding reference clip $\tilde{\tau}$ starting from $\tilde{\stt}_0$ in $\dataset$. This is achieved by recursively applying the control policy $\policy(\act_t|\stt_t,\latent_t)$ in the world model $\omega(\stt_{t+1}|\stt_t, \act_t)$.
In this process, the latent variables $\latent_t$ are sampled from the approximate posterior $q(\latent_t|{\stt}_t, \tilde{\stt}_{t+1})$, where the future states $\tilde{\stt}_{t+1}$ in the state transitions are extracted from  $\tilde{\tau}$. The reparameterization trick~\cite{Kingma2014_VAE} is used when sampling $\latent_t$ to ensure the backpropagation of the gradients.

Using all the approximations described above, we can now convert \eqn\eqref{eqn:ELBO} into loss functions. Considering \eqn\eqref{eqn:log_likelihood_transition}, we have
\begin{align}
    \loss_{\eqword{rec}} 
    &= \sum_{t=0}^{T-1} \gamma^t \E_{\begin{subarray}{l}q(\latent_t|{\stt}_{t},\tilde{\stt}_{t+1}),\policy(\act_t | \stt_t, \latent_t)\end{subarray}} [ \left\Vert \tilde{\stt}_{t+1} - \vect{\mu}_w(\stt_t,\act_t) \right\Vert^2_{W}  ]
    \label{eqn:rec_loss}
\end{align}
\begin{align}
    \loss_{\eqword{kl}} &= \sum_{t=0}^{T-1} \gamma^t {\left\Vert\vect{\mu}_q({\stt}_t, \tilde{\stt}_{t+1}) \right\Vert_2^2 }/{2{\sigma}_p^2}
    \label{eqn:kl_loss},
\end{align}
where the states labeled with a tilde ($~\tilde{}~$) are extracted from the reference clip  $\tilde{\tau}$ and those without it are from the synthetic trajectory. Considering that the synthetic trajectory is constructed using the approximate world model, the states in the trajectory can deviate gradually from the true simulation. A discount factor $\gamma$ is thus employed to lower the weights of these inaccurate states along the trajectory. Here we use $\gamma=0.95$ in this paper.

In addition to the above ELBO losses, following SuperTrack~\cite{fussellSuperTrackMotionTracking2021}, we regularize the magnitude of the actions in terms of both the $L_1$ and $L_2$ metrics to prevent excessive control. The corresponding loss term is defined as
\begin{equation}
    \mathcal{L}_{\text{act}} = \sum_{t=0}^{T-1} \gamma^t \left. \left(  w_{a_1}\left\|\act_t \right\|_1 + w_{a_2}\left\| \act_t \right\|^2_2 \right),\right.
\end{equation}
which is discounted similarly to the ELBO losses. 
In practice, we find $L_1$ term can also be replaced with a larger $w_{a_2}$ weight.

The final objective for training the \cvae{} is then given by
\begin{equation}
    \mathcal{L} = \mathcal{L}_{\text{rec}}  + \beta \mathcal{L}_{kl} +  \mathcal{L}_{\text{act}}.
    \label{eqn:loss_VAE}
\end{equation}
Following $\beta$-VAE~\cite{Higgins2017betaVAELB}, we employ a weight parameter $\beta$ for the KL-divergence loss, which increases once every $500$ training epochs from $0.01$ to $0.1$ during the training.

\subsubsection{World Model} 
\label{sec:training_world_model}
The learning process of the world model largely repeats the same process in SuperTrack~\cite{fussellSuperTrackMotionTracking2021}.
Specifically, $\omega(\stt_{t+1}|\stt_t, \act_t)$ is learned based on a collection of simulated trajectories, $\simBuffer=\{\tau^*_j\}$. We collect these trajectories by executing the current control policy in the simulation with the start states and skill latent variables extracted from random reference trajectories in $\dataset$.
We then solve an MLE problem again to train $\omega(\stt_{t+1}|\stt_t, \act_t)$. Similar to the learning of the \cvae{}, we generate a synthetic trajectory starting from a random state $\bar{\stt}_0$ in $\simBuffer$ by executing the recorded sequence of actions $\{\bar{\act}_t\}$ in the world model. The loss function is then computed as
\begin{align}
    \mathcal{L}_{w} = \sum_{t=0}^{T'-1} \left\Vert \bar{\stt}_{t+1} - \vect{\mu}_w(\stt_t, \act_t) \right\Vert^2_{W'},
    \label{eqn:loss_world_model}
\end{align}
where $\{\bar{\stt}_{t+1}\}$ are the recorded states corresponding to $\{\bar{\act}_t\}$.


\input{Sections/4.1_Pseudocode.tex}
\subsubsection{Training Process}
During training, we first collect a number of simulated trajectories then update the world model and the \cvae{} in tandem as illustrated in Algorithm~\ref{alg:training}. 

\paragraph{Trajectory collection.}
At the beginning of each training epoch, we select a start state $\stt_0$ from a random trajectory $\tilde{\tau}$ in the motion dataset $\dataset$. Then the control policy $\policy(\act_t|\stt_t,\latent_t)$ is evaluated to track $\tilde{\tau}$, using the latent skill variable sampled from the posterior distribution {$q(\latent_t|{\stt}_t,\tilde{\stt}_{t+1})$}, conditional on the current state {${\stt}_t$} of the character and the corresponding next reference state $\tilde{\stt}_{t+1}$ in $\tilde{\tau}$. The character then performs the action $\act_t$ in the simulation, resulting in a new state {${\stt}_{t+1}$}. This process is repeated until a termination condition is satisfied. We then store the simulated trajectory {${\tau}^*$ and corresponding reference trajectory $\tilde{\tau}$} in a temporary buffer $\simBuffer'$, and start a new trajectory from another random state. This trajectory collection procedure is ended when the number of states in $\simBuffer'$ exceeds a predefined size $N_{B'}$. Then the temporary buffer $\simBuffer'$ is merged into $\simBuffer$, replacing the oldest trajectories and keeping the size of $\simBuffer$ smaller than $N_B$. Here we use $N_B=5\times{}10^4$ and $N_{B'}=2048$.

We employ an early termination strategy to prevent $\simBuffer$ from recording too many bad simulation samples. A simulation will be terminated if the trajectory is longer than $T_{\eqword{max}}=512$ time steps or if the tracking error of the character's head has exceeded $d_{\eqword{max}}=0.5\,m$ for more than $T_{\eqword{term}}=1$ second.

\paragraph{Update world model.}
We then update the world model as described in \Sec\ref{sec:training_world_model}. More specifically, we extract a batch of $N_{\eqword{w}}$ random clips $\{\bar{\tau}^*\}$ of length $T_{\eqword{w}}$ from $\simBuffer$, where each $\bar{\tau}^*=\{ \bar{\stt}_{0:T_{\eqword{w}}},\bar{\act}_{0:T_{\eqword{w}}-1}\}$. Then $N_{\eqword{w}}$ synthetic trajectories are generated by unrolling the world model using $\bar{\stt}_0$ and $\{\bar{\act}_{0:T_{\eqword{w}}-1}\}$ to compute the loss in \eqn\eqref{eqn:loss_world_model}. The world model is then updated using the gradients of the loss function. This updating process is repeated $8$ times in each training epoch with $T_{\eqword{w}}=8$ and $N_{\eqword{w}}=512$.

\paragraph{Update \cvae{}} At last, the \cvae{} is updated as described in \Sec\ref{sec:training_control_VAE}. Similar to the training of the world model, we generate a batch of $N_{\eqword{VAE}}=512$ synthetic trajectories to evaluate the loss functions in \eqn\eqref{eqn:loss_VAE}, where each trajectory has $T_{\eqword{VAE}}=24$ frames. To ensure a good coverage over the state space, the start states of these synthetic trajectories are randomly selected from the simulation buffer $\simBuffer$, and the corresponding motion clips in $\dataset$ are extracted as the reference. We update the \cvae{} models $8$ times before starting the next training epoch.

\subsection{Implementation}

\subsubsection{Policy Representation}
\paragraph{State}
Our physics-based character is modeled as articulated rigid bodies with a floating root joint. Its state can be fully characterized by $\stt^*=\{\vect{x}_i, \vect{q}_i, \vect{v}_i, \vect{\omega}_i\}, i\in{}B$, where $B$ stands for the set of rigid bodies and $\vect{x}_i, \vect{q}_i, \vect{v}_i, \vect{\omega}_i$ are the position, orientation, linear velocity, and angular velocity of each rigid body, respectively. 
Similar to SuperTrack~\cite{fussellSuperTrackMotionTracking2021}, we convert the global state of the character $\stt^*$ into the local coordinate frame of the root, and use the result, $\stt$, as the input to the \cvae{} and the world model. Specifically, we define $\stt=\{ \bar{\vect{x}}_i, \bar{\vect{q}}_i, \bar{\vect{v}}_i, \bar{\vect{\omega}}_i, h_i, \vect{y}_0 \}, i\in{}B$, where
\begin{align}    
    (\bar{\vect{x}}_i, \bar{\vect{q}}_i, \bar{\vect{v}}_i, \bar{\vect{\omega}}_i) = 
    \vect{q}_0^{-1} \otimes (\vect{x}_i - \vect{x}_0, \vect{q}_i, \vect{v}_i, \vect{\omega}_i),
\end{align}
$h_i$ is the height of each rigid body, and $\vect{y}_0$ is the \emph{up} axis of the root's local coordinate frame. The rotations $\vect{q}$ are represented in 6D representation \cite{zhou2019continuity}, which are commonly used in recent research on motion synthesis \cite{fussellSuperTrackMotionTracking2021, leeLearningTimecriticalResponses2021}. They can be computed by extracting the first two columns of a rotation matrix.

With this state representation, the reconstruction loss of \eqn\eqref{eqn:rec_loss} is implemented in practice using the 1-norm distance between two states as
\begin{align}
    \loss_{\eqword{rec}} 
    &= \sum_{t=0}^{T-1} \gamma^t \left[ \Vert W(\tilde{\stt}_{t+1} - \stt_{t+1}) \Vert_1 \right], \label{eqn:loss_recon_impl}
\end{align}
where $W=\diag(w_{\bar{x}},w_{\bar{q}},w_{\bar{v}},w_{\bar{\omega}},w_h,w_y)$ is a diagonal weight matrix that balances the magnitude of each component.

\paragraph{Action}
We actuate our character using PD controllers. The action $\act=\{\vect{\hat{q}}_j\}, j\in{}J$ is thus a collection of target rotations of all the joints, where $J$ stands for the set of joints. We use 3D axis angles to represent those target rotations.

\paragraph{Latent}
We employ a latent space $\Latent$ with dimension $64$ to encode the motion skills. The means of the conditional prior $p(\latent_t|\stt_t)$ and the approximate posterior $q(\latent_t|\stt_t, \stt_{t+1})$ of the latent variable are both modeled as neural networks with two 512-unit hidden layers and the ELU as the activation function. To emphasize the state information, we concatenate the input of each layer of the prior and posterior networks with $\stt_t$ and $\stt_{t+1}$, respectively. We use $\sigma_{\vect{p}}=0.3$ as the standard deviation of these distributions.

\paragraph{Policy}
The mean $\vect{\mu}_{\pi}$ of the control policy $\policy(\act|\stt,\latent)$ is modeled as a neural network. We employ a mixture-of-expert (MoE) structure similar to the decoder network of MotionVAE~\cite{lingCharacterControllersUsing2020a}. Specifically, we use six expert networks in this structure, each having three 512-unit hidden layers. The parameters of these experts are blended according to the weights computed by a gating network, which contains two 64-unit hidden layers. The ELU is used as the activation function for these networks. To ensure the effectiveness of the latent skill variable, $\latent$ is concatenated with the input of each layer of the experts, and layer normalization is applied to normalize these concatenated inputs. The covariance matrix $\Sigma_{\pi}$ of the policy distribution is set as a diagonal matrix $\sigma_{\pi}^2\vect{I}$, where $\sigma_{\pi}=0.05$.

\subsubsection{World Model}
Our world model $\omega(\stt_{t+1}|\stt_t, \act_t) \sim \mathcal{N}(\vect{\mu}_w, \vect{\Sigma}_w)$ is formulated and trained in the same way as SuperTrack~\cite{fussellSuperTrackMotionTracking2021}. We use a neural network with four 512-unit hidden layers and ELU activation functions for the world model $\vect{\mu}_w(\stt, \act)$.
%
The input to $\vect{\mu}_w(\stt, \act)$ is the state representation $\stt$ and the target joint rotations of the action $\act$ converted into quaternions. The world model then predicts the change of the velocity and angular velocity of each rigid body, represented by $d \bar{\vect{v}}_i$ and $d \bar{\vect{\omega}}_i$, in the local coordinate frame of the root. Then, the character's new state $\stt^*_{t+1}$ and thus $\stt_{t+1}$ are computed accordingly using the forward Euler method.

 The objective function of \eqn\eqref{eqn:loss_world_model} is then implemented to penalize the prediction error in the global coordinate frame:
\begin{align}
    \mathcal{L}_{w} = \sum_{t=0}^{T'-1} \left\Vert \bar{\stt}^*_{t+1} - {\stt}^*_{t+1} \right\Vert^2_{W'},
\end{align}
where the weight matrix $W'=\diag(w_x,w_q,w_v,w_{\omega})$ is chosen empirically to balance the magnitude of each component.

Our experiments show that large randomness can negatively impact the stableness of the training process. In practice, we directly use the mean function  $\vect{\mu}_w$ to generate the synthetic rollouts instead of sampling from $\mathcal{N}(\vect{\mu}_w, \vect{\Sigma}_w)$.


\subsubsection{Data Balancing} 
Our motion dataset $\dataset$ consists of several unstructured long motion sequences of a diverse set of skills, including walking, running, jumping, turning, and more difficult skills such as getting up. During the trajectory collection process of the training, if we select the initial states uniformly from $\dataset$, easy motions may often result in longer simulated trajectories and thus occupy the simulation buffer $\simBuffer$, making the \cvae{} hard to learn difficult skills. To overcome this unbalancing problem, we calculate the \emph{value} of each state and sample the initial states according to their values. The state with a lower value will have higher chance to be selected. 

More specifically, we maintain a list $\mathcal{V}=\{V_t\}$ of the values of every state in the dataset $\dataset$, where all the $V_t$ are initialized to zero. After every 200 epochs in the training, if a state with index $t$ in $\dataset$ is encountered in the collected simulated trajectories, we update the corresponding value $V_t$ as
\begin{align}
    V_t^* &= \sum_{k=0}^{T-t}\gamma^{k}r_{t+k} + \gamma^{T+1-t}V_{T+1} \\
    V_t &= (1-\alpha)V_t + \alpha{}V_t^*
\end{align}
where $r_{t+k}$ is the reward of state $\stt_{t+k}$ in the simulated trajectory, $T$ is the length of the trajectory containing $\stt_t$, and the discount factor $\gamma=0.95$. We compute the reward according to the reconstruction loss of \eqn\eqref{eqn:loss_recon_impl} as
\begin{align}
    r_t = e^{-{\Vert W(\tilde{\stt}_{t} - \stt_{t}) \Vert_1}/{T_V}}, \label{equ:reward}
\end{align}
where the temperature $T_V=20$. In the next 200 training epochs, the initial state will be selected based on the probability proportional to~$1/\max(0.01,V_t)$.


\subsubsection{Trajectory Sampler}
\label{sec:trajectory_sampler}
At last, to encourage the policy to learn new transitions between different motions, we augment the trajectory collection process by switching the reference motion clip randomly during the simulation. The beginning state of the new motion clip is also selected using the above data balancing strategy. In practice, many of such switches will cause the character to fall due to mismatched poses and velocities, but the character can learn to recover from such falls automatically during the training.

To encourage the policy to embed diverse motion transitions into prior distribution $p(\latent|\stt)$, we further augment the trajectory collection process with states derived from the prior distribution. Specifically, the policy randomly chooses to sample latent codes from the posterior distribution $q(\latent_t|\stt_t,\tilde{\stt}_{t+1})$ or the prior distribution $p(\latent_t|\stt_t)$. The probability of choosing the latter is empirically set to $0.4$. In practice, the skill embeddings learned with this augmentation behave more actively and responsively in downstream tasks.


%% file: Sections/4.1_Pseudocode.tex

\begin{algorithm}[t] 
    
  \SetAlgoLined
  
  \SetKwProg{TrajectoryCollection}{Function}{:}{end}
  \TrajectoryCollection{ \textnormal{\textbf{TrajectoryCollection}$($ \textbf{Env}, \textbf{ControlVAE}$)$} }
  {
    Select $\tilde{\tau}=\{\tilde{\stt}_0,\tilde{\stt}_1,\dots{},\tilde{\stt}_T\}$ from \dataset \;
    ${\stt}_0 \leftarrow \tilde{\stt}_0$, $t \gets 0$\;
    
    \While{\textnormal{not terminated}}{
        \tcc{Get action from ControlVAE}
        Sample $\latent_t \sim{} q(\latent_t|{\stt}_t,\tilde{\stt}_{t+1})$\;
        \tcp{or $\latent_t \sim{} p(\latent_t|{\stt}_t)$, see Section~\ref{sec:trajectory_sampler}}
        Sample $\act_t \sim{} \policy(\act_t|{\stt}_t,\latent_t)$ \;
        ${\stt}_{t+1} \leftarrow $ simulation with state ${\stt}_t$ and action $\act_t$\;
        
        $t \leftarrow t+1$\;
    }
    Store ${\tau}^*=\{{\stt}_0, \act_0, {\stt}_1, \act_1, \dots, \}$ and $\tilde{\tau}=\{\tilde{\stt}_0, \tilde{\stt}_1, \dots \}$in $\simBuffer'$\;
  }
  
  \SetKwProg{TrainWorldModel}{Function}{:}{end}
  \TrainWorldModel{ \textnormal{\textbf{TrainWorldModel}$( \omega$, $T_{\eqword{w}})$  }}
  {
    Sample $\bar{\tau}^*=\{\bar{\stt}_0, \bar{\act}_0, \bar{\stt}_1, \bar{\act}_1, \dots, \}$ from $\simBuffer$\;
    $\stt_0 \leftarrow \bar{\stt}_0$, $\loss{}_{\eqword{w}} \gets 0$\; 
    
    \tcc{Generate synthetic trajectories}
    \For{$t\gets0$ \KwTo $T_{\eqword{w}}-1$}{
        $\stt_{t+1} \gets \omega(\stt_{t},\bar{\act}_{t})$\;
        $\loss{}_{\eqword{w}} \gets \loss{}_{\eqword{w}} + \|\stt_{t+1}-\bar{\stt}_{t+1} \|_{W'}$
    }
    Update $\omega$ with $\loss{}_{\eqword{w}}$
  }

  \SetKwProg{TrainControlVAE}{Function}{:}{end}
  \TrainWorldModel{ \textnormal{\textbf{TrainControlVAE}$($ $\omega$, \textbf{ControlVAE}, $T_{\eqword{\eqword{VAE}}}$ $)$} }
  { 
    Sample $\bar{\tau}^*$ and $\tilde{\tau}$ from $\simBuffer$\;
    $\stt_0 \leftarrow \bar{\stt}_0$, $\mathcal{L} \gets 0$\;
    
    \tcc{Generate synthetic trajectories}
    
    \For{$t\gets0$ \KwTo $T_{\eqword{VAE}}-1$}{
        Sample $\latent_t \sim{} q(\latent_t|{\stt}_t,\tilde{\stt}_{t+1})$\;
        Sample $\act_t \sim{} \policy(\act_t|\stt_t,\latent_t)$ \;
        $\stt_{t+1} \gets \omega(\stt_{t},\act_{t})$\;
        $\loss \gets \loss + \loss_{\text{rec}}(\stt_{t+1},\tilde{\stt}_{t+1})  + \beta \loss_{kl} +  \loss_{\text{act}}$
    }
    Update \textbf{\cvae{}} with $\loss$
  }
  
  \caption{Training algorithm of \cvae{}}
  \label{alg:training}
\end{algorithm}

%% file: Sections/5_High_Level_Control.tex
\section{Model-Based High-Level Controllers}
\label{sec:high-level}

A learned \cvae{} provides a latent skill space $\Latent$ and a skill-conditional policy $\policy(\act|\stt,\latent)$ that a high-level policy can leverage to accomplish various downstream tasks. Formally, the task policy $\policy(\latent|\stt,\task)$ takes the character state $\stt$ and a task-specific parameter~$\task$ as input and computes a skill variable $\latent$. The skill policy $\policy(\act|\stt,\latent)$ then uses $\latent$ to compute the action $\act$ accordingly.
The task policy can be trained using model-free reinforcement learning algorithms as suggested in previous systems \cite{pengMCPLearningComposable2019,luoCARLControllableAgent2020,merelCatchCarryReusable2020,lingCharacterControllersUsing2020a}. 
%
%
However, the world model learned with \cvae{} further enables more efficient model-based learning for the downstream tasks.
%

In this section, we introduce two model-based control strategies that can take advantage of the learned world model, as well as a corpus of locomotion tasks that can be accomplished using these controllers. Note that the parameters of the learned \cvae{} are frozen in this stage. Only the gradients are passed through the networks to optimize the downstream task policies. The character applies the learned policies in the true simulation and responds dynamically to user control and unexpected perturbations.






\subsection{Model Predictive Control}
The first control strategy is the sampling-based model predictive control (MPC).  At each time step, we generate $N_{\eqword{MPC}}$ synthetic Monte-Carlo rollouts $\{\tau_i^{\task}=\{\latent_t,\act_t,\stt_t\}^{i}\}$ of a fixed planning horizon $T_{\eqword{MPC}}$ using the world model, 
where the skill variables $\latent_t$ of these rollouts are sampled from the state-conditional prior distribution $p(\latent_t|\stt_t)$ of the \cvae{}. We then evaluate each rollout with a task-specific loss function. The first action of the best trajectory will be used in the simulation. In practice, we find that $N_{\eqword{MPC}}=128$ and $T_{\eqword{MPC}}=4$ can achieve good performance in our experiments.


\subsection{Model-based Learning of Task Policies}
\label{sec:supervised_learning}


Our \cvae{} allows fast model-based policy training for downstream tasks based on the world model. Specifically, we assume that the task policy $\policy(\latent_t|\stt_t,\task)$ has the same structure as the approximate posterior distribution $q(\latent_t|\stt_t,\stt_{t+1})$ of \cvae{}, which is again a Gaussian distribution $\mathcal{N}(\hat{\vect{\mu}}_{g},\sigma_{g}^2\vect{I})$ with a diagonal covariance $\sigma_{g}=\sigma_q$ and the mean function computed as 
\begin{align}
    \hat{\vect{\mu}}_{g}=\vect{\mu}_p + \vect{\mu}_{g},
\end{align}
where $\vect{\mu}_{g}=\vect{\mu}_{g}(\stt,\task;\theta_g)$ is a neural network parameterized by $\theta_g$. During the training, we generate a batch of $N_{\eqword{ML}}=256$ synthetic rollouts $\{\tau^{\task}_i\}$ for a fixed horizon $T_{\eqword{ML}}=16$ and update the policy $\vect{\mu}_{g}$ by minimizing the task-specific loss function. The reparameterization trick is also applied in this process to ensure the backpropagation of the gradients to the policy. 

To generate the initial states for synthesizing these rollouts, we apply the current task policy in the simulated environment with the corresponding goal parameters $\task$ changing randomly every 72 steps. These simulated states and goal parameters are then stored in a buffer $\simBuffer_{\task}$ with size $4096$. During the training, a synthetic rollout is generated by sampling an initial state $\stt_0$ from $\simBuffer_{\task}$ and executing the task policy using the recorded sequence of task parameters $\{\task_t\}$ following $\stt_0$.



\subsection{Tasks}
In this section, we introduce a corpus of locomotion tasks that can be accomplished by the above model-based control strategies. We formulate these tasks as a set of loss functions computed over each of the generated trajectories $\tau^{\task}$. 
Using the skill embeddings learned in \cvae{}, the corresponding task policies can generate natural motions to complete these tasks in simulation and respond to unexpected perturbations.

Unless otherwise stated, all our loss functions have the form
\begin{align}
    \mathcal{L}(\tau^{\task})=
    \sum_{t=1}^{T} \left[\mathcal{L}_{\task}(\stt_{t-1}, \stt_t) + \mathcal{L}_{\eqword{fall}}(\stt_t) \right]
    + w_{z}\sum_{t=0}^{T-1} \Vert \vect{\mu}_{g} \Vert_2^2 ,
    \label{eqn:high_level_policy_loss}
\end{align}
where $\mathcal{L}_{\task}$ stands for the task-specific loss computed using the current state $\stt_t$ and optionally the previous state $\stt_{t-1}$, and $\mathcal{L}_{\eqword{fall}}(\stt_t)$ penalizes potential falling as
\begin{align}
    \mathcal{L}_{\eqword{fall}}(\stt_t)=\max(h_0^*-h_0,0),
\end{align}
where $h_0$ is the height of the character's root, $h_0^*=0.5\,m$ is a falling threshold. The last term of \eqn\eqref{eqn:high_level_policy_loss} is a regularization term that encourages the high-level policy to stay close to the prior distribution. We find this term crucial to a stable and successful training. $w_z=20$ works for all our experiments. 

\paragraph{Height control}
We define a simple loss to control the height of the character as

\begin{align}
    \mathcal{L}_{\task} &= H \cdot h_0 ,
\end{align}
where $h_0$ represents the current height of the character's root joint. The task parameter $H\in\{-1,1\}$ indicates the desired motion for the character, where $H=-1$ encourages the character to squat down and eventually lie on the ground and $H=1$ will let the character get up and jump when possible to maintain a high root position. 

\paragraph{Heading control} In the heading control task, the character needs to travel while heading towards a target direction $\theta_h^*\in{}[-\pi,\pi]$ at a given speed $v^*\in{}[0.0,3.0]\,m/s$ along that direction. The loss function is thus defined as 
\begin{align}
    \mathcal{L}_{\task} &= w_{\theta_h} | \theta_h^* - \theta_h | + w_{v} \frac{|v^*- v|}{\max(v^*, 1)},
    \label{eqn:heading_loss}
\end{align}
where $\theta_h$ and $v$ are the character's current heading direction and speed. We compute $\theta_h$ according to the orientation of the character's root and $v$ as the component of the root's velocity along this direction. The weights are set to $(w_{\theta_h}, w_{v})=(2.0, 1.0)$ for this task. Note that we normalize the speed loss according to the target speed to encourage the training to pay attention to these low-speed motions.


\paragraph{Steering control} The objective of the steering task is to control both the heading direction $\theta_h$ and the travel direction $\theta_v$ simultaneously. Specifically, we define the loss function as
\begin{align}
    \mathcal{L}_{\task} &= 
    w_{\theta_h} | \theta_h^* - \theta_h | +
    w_{\theta_v} | \theta_v^* - \theta_v | +
    w_{v} \frac{|v^*- \| \bar{\vect{v}}_0 \|_2|}{\max(v^*, 1)},
\end{align}
where $\bar{\vect{v}}_0$ represents planar components of the root's linear velocity $\bm{v}_0$ and $\theta_v$ is its directional angle, computed as the angle between $\bar{\vect{v}}_0$ and the x-axis of the global coordinate frame.
The weights are set similarly as $(w_{\theta_h}, w_{\theta_v}, w_{v})=(2.0,2.0,1.0)$ for this task.

\input{Sections/_Algorithm_Skill}
\paragraph{Skill control} 
The objective of the skill control is to let the character use a specific skill to accomplish a given task, such as responding to steering control while jumping or hopping. In this task, a skill is specified using a 4-second motion clip manually selected from the dataset. We have selected $5$ skills for this task, including walking, running, hopping, jumping, and skipping, as shown in \fig\ref{fig:simulated_skills}. The task parameter $\vect{c}$ is thus defined as a one-hot vector indicating the index of the skill.

Unfortunately, we do not yet have a direct mapping between a specific skill $\vect{c}$ and its corresponding latent code $\latent$, so that the character has to figure out by itself whether it is performing the correct skill corresponding to the reference motion clip. Inspired by AMP~\cite{pengAMPAdversarialMotion2021a}, which learns an adversarial discriminator to enforce a specific motion style in reinforcement learning, we employ a similar discriminator in our model-based learning process that penalizes incorrect motions as an adversarial loss function.



More specifically, we train the task-specific control policy to mimic the behavior of a tracking policy characterized by the approximate posterior $q(\latent_t|\stt_t, {\stt}_{t+1})$, as illustrated in Algorithm~\ref{alg:skill_task}. In each training iteration, we first generate a batch of control rollouts $\{\tau^{\task}\}$ as described in \Sec\ref{sec:supervised_learning}, where each $\tau^{\task}$ is conditioned on a skill vector $\vect{c}$. Then, for each $\tau^{\task}$, we extract a short random reference clip with the same length from the reference motion of skill $\vect{c}$. By tracking this short reference clip using the approximate posterior $q(\latent_t|\stt_t, \tilde{\stt}_{t+1})$, where $\stt_t$ is the generated state and $\tilde{\stt}_{t+1}$ is from the reference clip, we obtain a tracking rollout $\tau^{\vect{c}}$. We then consider $\{\tau^{\vect{c}}\}$ as the \emph{real} data samples and the control rollouts $\{\tau^{\task}\}$ as the \emph{fake} data samples and employ a discriminator $D$ to distinguish between them.
Following \citet{pengAMPAdversarialMotion2021a}, we adapt a least-squares GAN \cite{Mao_LSGAN_YHY} in this training. The adversarial loss for the discriminator $D$ is defined as
\begin{align}
    \argmin_D &\ \E_{\stt_t, \stt_{t+1}\sim \{\tau^{\vect{c}}\}} \left[ (D(\stt_t, \stt_{t+1};\vect{c})-1)^2\right] \nonumber\\
    &+ \E_{\stt_t, \stt_{t+1}\sim \{\tau^{\task}\}} \left[ (D(\stt_t, \stt_{t+1};\vect{c})+1)^2\right] \nonumber\\
    &+ w_g \E_{\stt_t, \stt_{t+1}\sim \{\tau^{\vect{c}}\}}\left[ \| \nabla_{\stt_t, \stt_{t+1}}  D(\stt_t, \stt_{t+1};\vect{c})  \|^2 \right].
    \label{eq:discriminator_loss}
\end{align}
Note that the discriminator $D$ is also conditioned on the skill vector~$\vect{c}$. The last term of the above equation regularizes the gradient of the discriminator with respect to the real data samples, which allows a more stable training~\cite{pengAMPAdversarialMotion2021a}. We set its weight $w_g = 20$.

The objective of the character is now to complete a target task, \eg{} heading control, while behaving indistinguishably from the reference motion. This can be achieved using the adversarial loss
\begin{align}
    \mathcal{L}_D= \left(D(\stt_t, \stt_{t+1};\vect{c})-1\right)^2,
\end{align}
where $\stt_t, \stt_{t+1}\sim \{\tau^{\task}\}$.

To further facilitate the character to perform the correct skill in the skill set, we train a separate skill classifier ${C}$ to predict the possiblity that a state transition $(\stt_t, \stt_{t+1})$ belongs to a specific skill~$\vect{c}$. It is trained with the transitions in the \emph{real} data samples $\{\tau^{\vect{c}}\}$ 
\begin{align}
    \argmin_C \E_{\stt_t, \stt_{t+1}\sim \{\tau^{\vect{c}}\}} \left[
        \mathcal{H}\left[ \mathcal{C}( \stt_t, \stt_{t+1}), c\right]
        \right],
\end{align}
where $\mathcal{H}$ is the {cross-entropy} loss. Then, we add a classifier loss 
\begin{align}
    \mathcal{L}_{C} = 
        \mathcal{H}\left[ \mathcal{C}( \stt_t, \stt_{t+1}), c\right]
        \label{eq:classifier_loss}
\end{align}
to the task's loss function, where again $\stt_t, \stt_{t+1}\sim \{\tau^{\task}\}$.

\begin{figure}[t]
    \centering
    \begin{subfigure}{0.35\linewidth}
      \includegraphics[width=\linewidth]{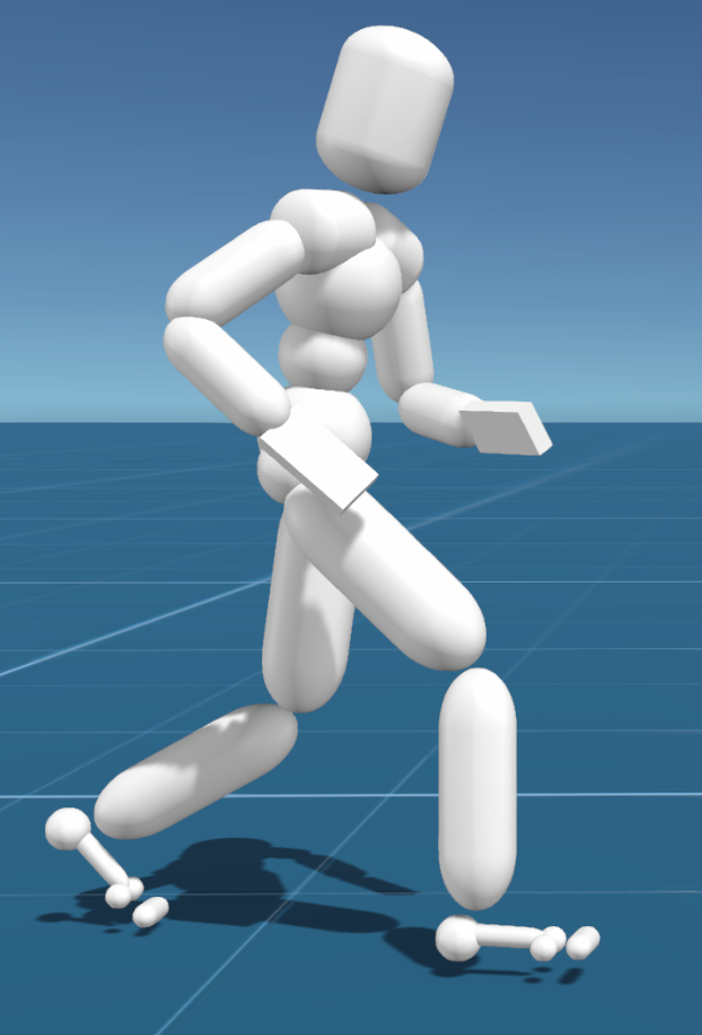}
      \caption{Simulation Model}
    \end{subfigure} \hspace{1.5mm}
    \begin{subfigure}{0.35\linewidth}
      \includegraphics[width=\linewidth]{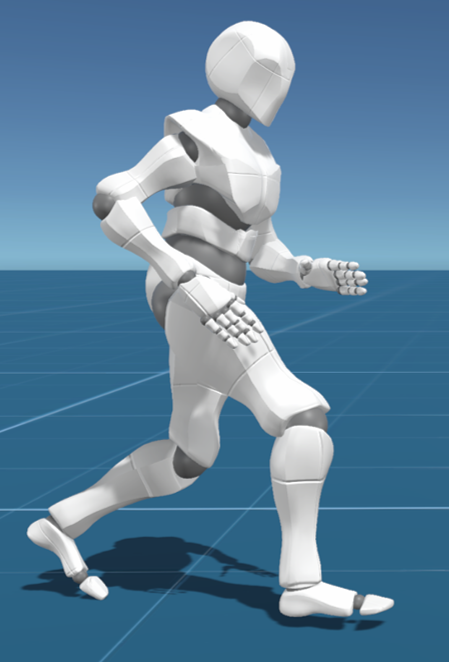}
      \caption{Visualization Model}
    \end{subfigure}
    \Description{}
    \caption{The character model used in this paper.}
    \label{fig:character_model}
\end{figure}

The final loss function for the style control is defined as
\begin{equation}
    \mathcal{L}_{\task} = \mathcal{L}_{\task'} + w_D\mathcal{L}_D + w_C\mathcal{L}_C + w_r\mathcal{L}_{\eqword{reg}},
    \label{eq:skill_loss}
\end{equation}
where $\mathcal{L}_{\task'}$ is the loss function of the target task. 
The last term of \eqn\eqref{eq:skill_loss}, $\mathcal{L}_{\eqword{reg}}$, is a regularization term inspired by \cite{luoCARLControllableAgent2020}. It considers the result of synthetically tracking the reference clip ${\tau}^{\vect{c}}$ of skill $\vect{c}$ as the output of a perfect policy finishing the task specified by ${\tau}^{\vect{c}}$ itself. The task policy being trained thus need to clone the behavior of this perfect policy. To achieve this, we extract the task parameters $\{{\task}^{\vect{c}'}_t\}$ from~${\tau}^{\vect{c}}$ and assume that the posterior ${\latent}^{\vect{c}}_t\sim{}q(\latent_t|{\stt}_t, \tilde{\stt}_{t+1})$ is the output of the perfect policy. 
The regularization term is then defined as 
\begin{equation}
    \mathcal{L}_{\eqword{reg}} = \|{\latent}^{\vect{c}}_t - \latent^{\vect{c}'}_{t}\|_1,
\end{equation}
where $\latent^{\vect{c}'}_{t} \sim \pi(\latent|\stt_t, {\task}^{\vect{c}'}_t)$ is the prediction of the current task policy for the task.
This term is crucial to maintain a stable training in practice. We use $(w_D, w_C, w_r)=(5.0, 0.5, 10.0)$ in our experiments.



%% file: Sections/_Algorithm_Skill.tex

\begin{algorithm}[t] 
  \SetAlgoLined

    \SetKwProg{GenSynTraj}{Function}{:}{end}
  \GenSynTraj{ \textnormal{\textbf{GenSynTraj}( \textbf{$\omega$,$\stt_0$, $p(\latent|\stt_t, *)$ ) } }}
  {
  \tcp{$p(\latent|\stt_t, *)$ is a conditional distribution of $\latent$}
    \For{$t\gets0$ \KwTo $T_{\eqword{SL}}-1$}{
        
        Sample $\latent_t \sim{} p(\latent|\stt_t, *)$\;
        Sample $\act_t \sim{} \policy(\act_t|\stt_t,\latent_t)$ \;
        $\stt_{t+1} \gets \omega(\stt_{t},\act_{t})$\;
    }
  }
  
  \SetKwProg{TrainSkillControl}{Function}{:}{end}
  \TrainSkillControl{ \textnormal{\textbf{TrainSkillCtrl}( \textbf{Env}, \textbf{ControlVAE}, $\omega$,  $\pi, D,C$ ) } }
  {
    \tcp{$D$ is the discriminator and $C$ is the classifier}
  
    Collect simulated trajectory $\bar{\tau}^*$ with random goals $\{\task_t\}$ \;
    Store $\bar{\tau}^*$ and $\{\task_t\}$ into $\simBuffer$\;
    
    
    \BlankLine
    \tcc{Synthesize \textit{real} rollouts}
    Select motion clips $\tilde{\tau}$ with a random skill label $\vect{c}$ in \dataset \;
    
    $\tau^{\vect{c}} = \{\stt^{\vect{c}}_0, \latent^{\vect{c}}_0 \dots\} \leftarrow$  \textbf{GenSynTraj}( $\tilde{\stt}_0$, $q(\latent|\stt_t,\tilde{\stt}_{t+1})$) \;

    \BlankLine
    \tcc{Synthesize fake rollouts}
    Select $\bar{\tau}^*$ and $\{\task_t\}$ from \simBuffer\;
    $\tau^{\task} = \{\stt_0, \stt_1, \dots \} \leftarrow$ \textbf{GenSynTraj}( $\bar{\stt}_0$, $\pi(\latent|\stt_t, \task_t)$) \;
    Calculate $\mathcal{L}_{\task'},\mathcal{L}_D, \mathcal{L}_C$ using $\tau^{\task}$ \;
    
    \BlankLine
    \tcc{Regularization term}

    \For{$t\gets0$ \KwTo $T_{\eqword{SL}}-1$}{
        $\task^{\vect{c}'}_t \leftarrow $ calculate goal parameters from $\stt^{\vect{c}}_t, \stt^{\vect{c}}_{t+1}$ \;
        $\latent^{\vect{c}'}_t \sim \pi(\latent|\stt^{\vect{c}}_t,\task^{\vect{c}'}_t)$ \;
        Calculate $\loss_{\eqword{reg}}$ using $\latent^{\vect{c}'}_t$, $\latent^{\vect{c}}_t$ \;
    }
    
    
    
    \BlankLine
    \tcc{Update network parameters}
    Update $\pi$\ with \eqn \eqref{eq:skill_loss}\;
    Update $D$ using $\tau^{c}, \tau^\eqword{fake}$ with \eqn \eqref{eq:discriminator_loss}\;
    Update ${C}$ using $\tilde{\tau}, \vect{c}$ with  \eqn \eqref{eq:classifier_loss}\;
  }
  \caption{Model-based training algorithm of task policy}
  \label{alg:skill_task}
\end{algorithm}

%% file: Sections/7_Results.tex
\section{Results}

\subsection{System Setup}
As shown in \fig\ref{fig:character_model}, we simulate a character model that is $1.6$\,m tall, weighs $49.5$\,kg, and consists of 20 rigid bodies. A uniform set of PD control parameters $(k_p, k_d) = (400,50)$ is used for all the joints, except for the toe joints $(10,1)$ and the wrist joints $(5,1)$. The character is simulated using the Open Dynamics Engine (ODE) at $120$\,Hz. The stable-PD mechanism~\cite{tanStableProportionalDerivativeControllers2011b,Liu2013_Simulation} is implemented to ensure a stable simulation with the large timestep.

Our system executes \cvae{}, the world model, and all the task-specific policies at 20\,Hz. It is lower than the simulation frequency, thus the same action is used in the simulation until the control policies are evaluated next time. \cvae{} assumes no knowledge about the simulation. It extracts the position and orientation of each rigid body from the physics engine and computes the corresponding velocities and angular velocities using finite difference.
We implement and train the \cvae{} using PyTorch~\cite{pytorch2019}. 
Once trained, the entire system runs faster than real-time on a desktop, allowing interactive control of the simulated character.

\begin{table}[t]
  \centering
  \caption{Length of each motion after resampling}
  \label{tab:motionlength}
  \begin{tabular}{|c|c|}
  \hline
  Motion & Frames (20\,fps) \\ \hline
  Walk  & 5227 \\ \hline
  Run   & 4757\\ \hline
  Jump  &  4889\\ \hline
  Getup &  3365\\ \hline
  \end{tabular}
\end{table}
\subsection{ControlVAE training}
We train our ControlVAE on an unstructured motion capture dataset consisting of four long motion sequences selected from the open-source LaFAN dataset \cite{harvey2020robustLafanYHY} as shown in \Tab\ref{tab:motionlength}. This dataset contains a diverse range of motions, including walking, running, turning, hopping, jumping, skipping, falling, and getting up. The motion sequences are downsampled to 20\,fps and augmented with their mirrored sequences, resulting in an augmented dataset with approximately 30\,minutes high-quality motion.

\begin{figure}[t]
  \centering 
  \includegraphics[width=0.9\linewidth]{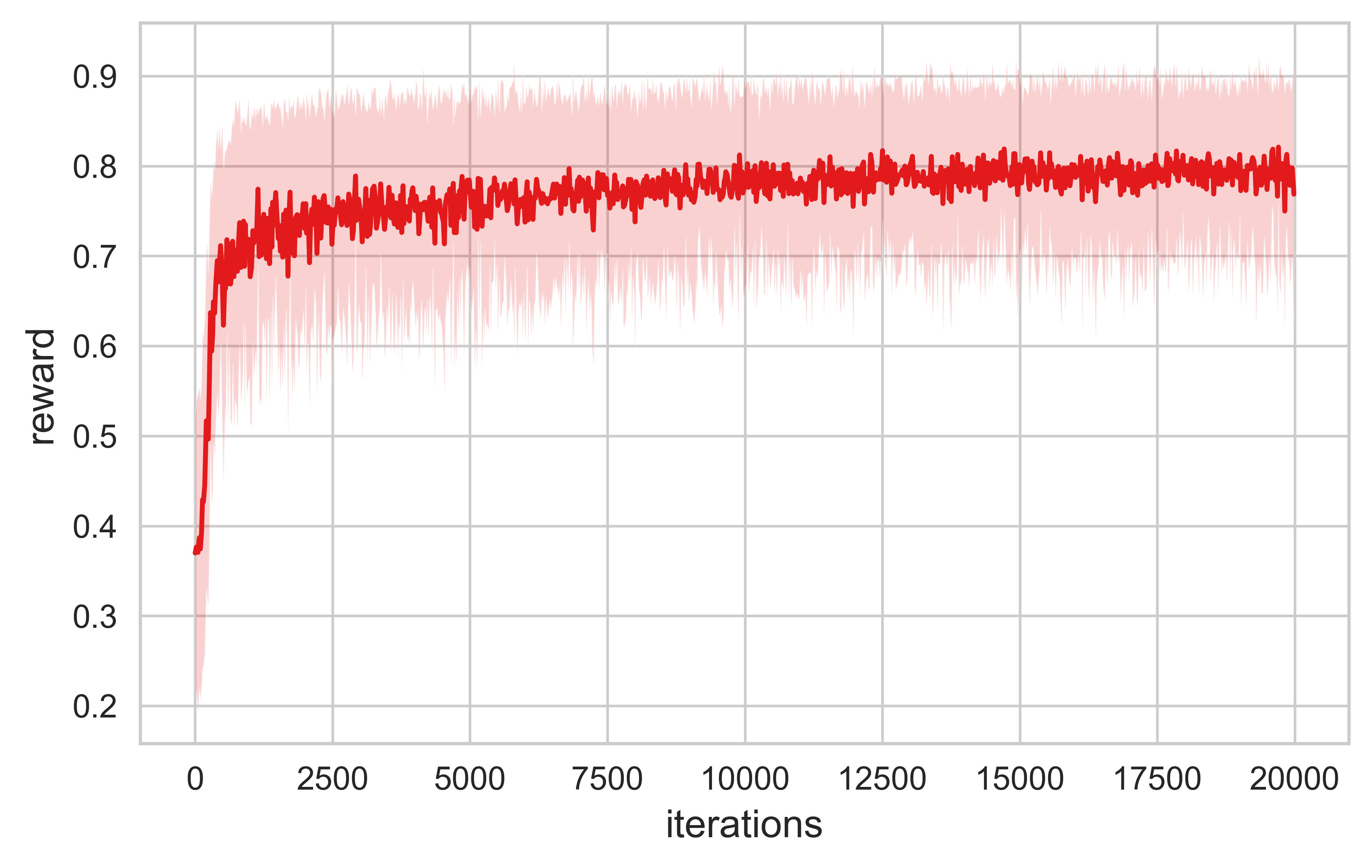}
  \caption{A typical learning curve of \cvae{}.}
  \label{fig:learning_curve_cvae}
\end{figure}

We train the \cvae{} models using the RAdam~\cite{liu2019radam} optimizer with $\beta_1 = 0.9$, $\beta_2 = 0.999$, and the learning rates of policy and world model are (1e-5, 2e-3). Following the standard technique for achieving a robust training, the gradient norms are clipped between $[-1,1]$. \fig\ref{fig:learning_curve_cvae} shows a typical learning curve of \cvae{}, where the reward is calculated using \eqn\ref{equ:reward} on the simulated trajectories collected during the training. The training process begins to converge after about 10,000 iterations, and the motion quality keeps improving in the following training. Our full training takes 20,000 iterations and about 50 hours with {four} parallel working threads on an {Intel Xeon Gold 6240 @ 2.60GHz} CPU and {one} NVIDIA  GTX 2080Ti graphics card. 

\subsection{Evaluation}
We evaluate the effectiveness of a learned \cvae{} using two simple tasks.

\paragraph{Reconstruction} 
To show that the \cvae{} has the same ability as other autoencoders in reconstructing an input motion, we use the learned approximate posterior $q(\latent_t|\stt_t,\stt_{t+1})$ as a tracking control and let \cvae{} to track motion sequences randomly chosen from the dataset. The character performs the input motion accurately in the simulation, and when another random clip is given, it automatically performs a smooth transition and then tracks the target motion again. In some extreme cases where the difference between current state and target motion is too large, the character may fall on the ground. It then automatically discovers a recovery strategy to get up, which is not included in the dataset. 



\paragraph{Random sampling.}

\begin{figure}[t]
  \centering
  \includegraphics[width=\linewidth]{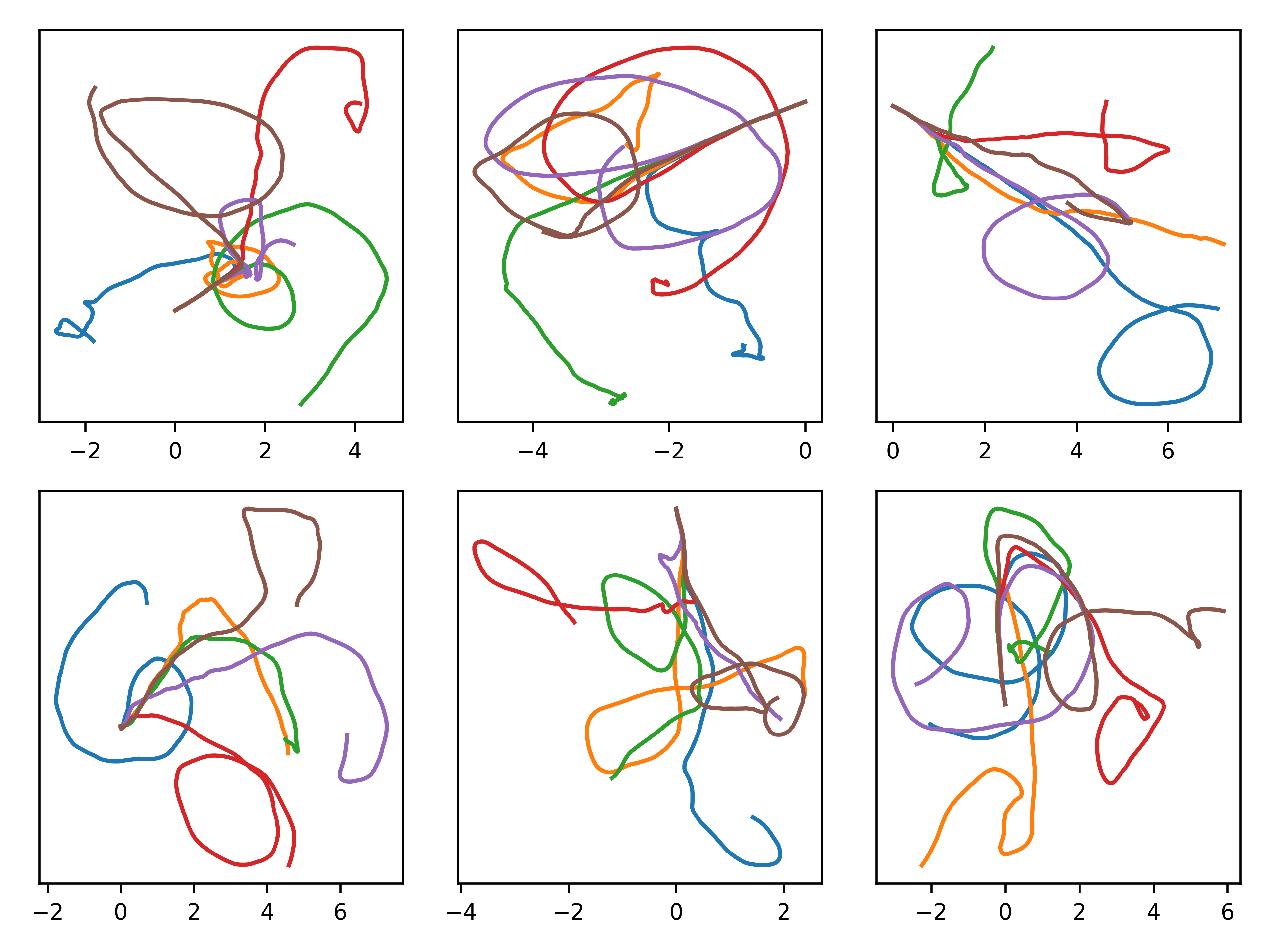}
  \caption{Visualization of the random trajectories generated in the random sampling experiment, showing the planar positions of the character. These trajectories are generated by drawing random latent codes from the prior distribution of a learned ControlVAE model. Starting from the same initial states, the simulated character moves towards different directions while performing realistic motions.}  
  \Description{}
  \label{Fig:randomwalk}
\end{figure}

The \cvae{} learns a generative control policy $\policy(\act|\stt,\latent)$ conditional on the latent skill variable $\latent$, which allows us to generate a diverse range of behaviors even by drawing random samples in the latent space $\Latent$. Specifically, given a random initial state, we draw samples from the state-conditional prior distribution $p(\latent|\stt)$ at every control step. The simulated character then performs random skills smoothly, such as stopping and then starting to walk, taking random turns, hopping and skipping for a short period, etc. \fig\ref{Fig:randomwalk} shows the root trajectories of several sample trajectories, where the character starts from a random state and performs random actions for 200 control steps, or 10 seconds. It can be seen that our \cvae{} generates a diverse range of motions in this random walk test.




\subsection{Model Predictive Control}
We test our MPC strategy on the height and heading control tasks. With interactive user input, the character lies down, gets up, and moves toward the target direction. 
The MPC policy can achieve real-time performance on the multicore computer we used to train \cvae{} with an NVIDIA GTX 2080Ti graphics card. Due to the limited number of samples, our sampling-based MPC cannot guarantee smooth and accurate results, but the resulting motions are generally acceptable. In practice, MPC can be used as an experimental tool to test our training settings, for example, to see whether a loss function is suitable for the task.

\subsection{Model-based Training}
\paragraph{Training settings} 
We model the task policy $\vect{\mu}_{g}$ using a neural network with three hidden layers, each having 256 units for heading control task and 512 for all other tasks. We use the RAdam~\cite{liu2019radam} optimizer with the learning rate decaying exponentially as 
$0.001*\max (0.99^{\eqword{iteration}}, 0.1)$.
We find this exponentially decaying learning rate improves the stability of the training process. 

\begin{figure}[t]
  \centering
    \includegraphics[width=0.7\linewidth]{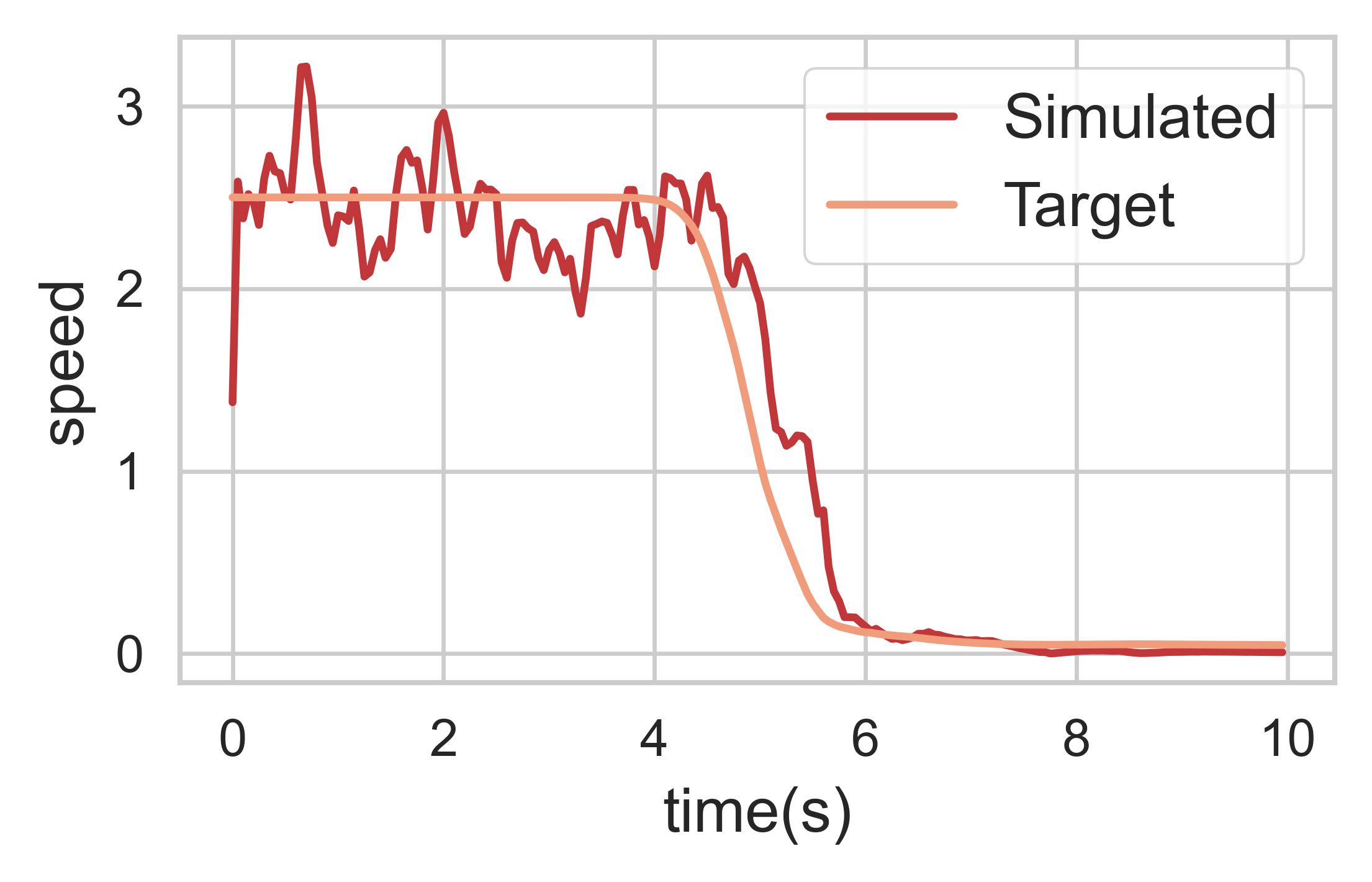}
  \caption{Visualization of the character's speed in the heading control task. The character slows down and stops under user control.}
  \label{fig:target_speed}
\end{figure}

\begin{figure*}[t]
  \centering
 
  \begin{subfigure}{0.965\linewidth}
    \includegraphics[width=\linewidth]{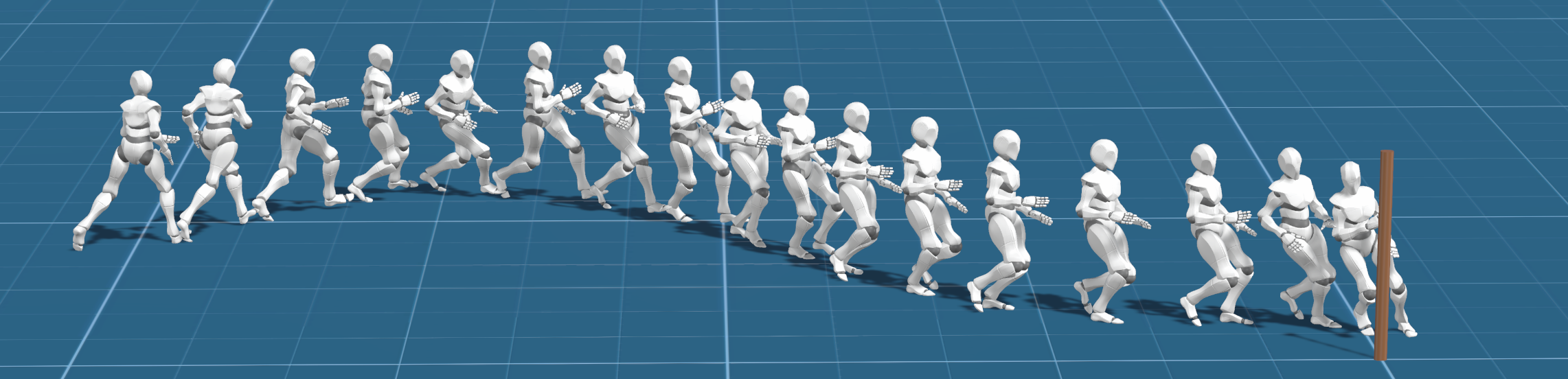}
  \caption{Target Location}
  \label{Fig:target}
  \end{subfigure}
  
   \begin{subfigure}{0.48\linewidth}
  \includegraphics[width=\linewidth]{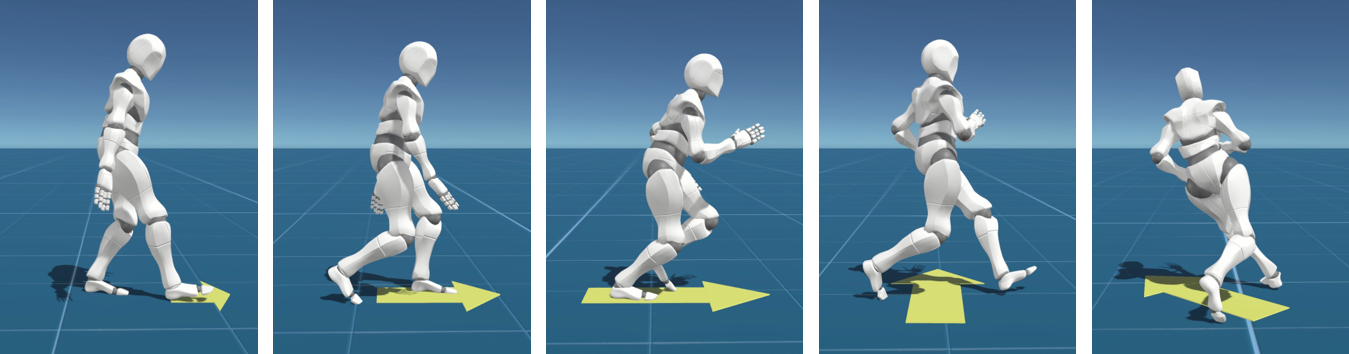}
  \caption{Speed and Turn}
  \label{Fig:speed}
  \end{subfigure}\hspace{1mm}
   \begin{subfigure}{0.48\linewidth}
  \includegraphics[width=\linewidth]{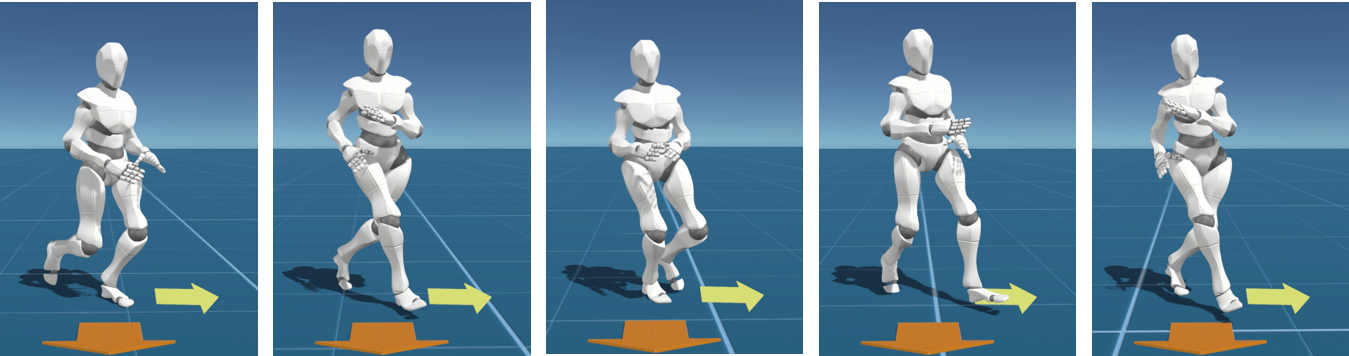}
  \caption{Walk Sideways}
  \label{Fig:walk_aside}
  \end{subfigure}
  
  \begin{subfigure}{0.48\linewidth}
  \includegraphics[width=\linewidth]{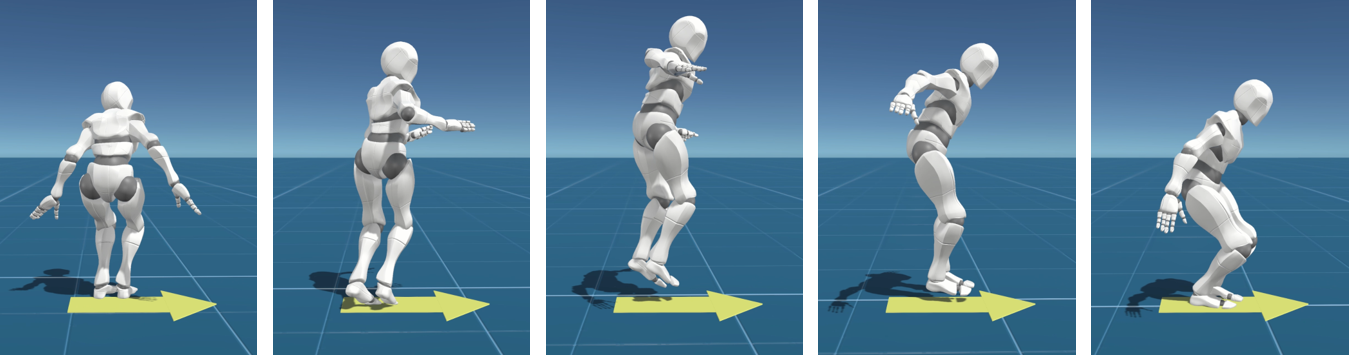}
  \caption{Jump and turn}
  \label{Fig:Jump_turn}
  \end{subfigure}\hspace{1mm}
  \begin{subfigure}{0.48\linewidth}
  \includegraphics[width=\linewidth]{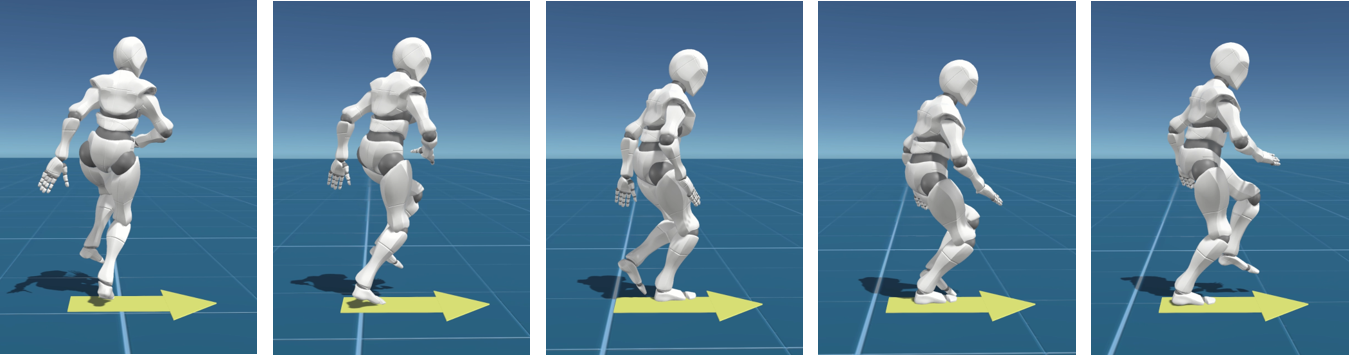}
  \caption{Hop and Turn}
  \label{Fig:hop_turn}
  \end{subfigure}
  
  \begin{subfigure}{0.48\linewidth}
    \includegraphics[width=\linewidth]{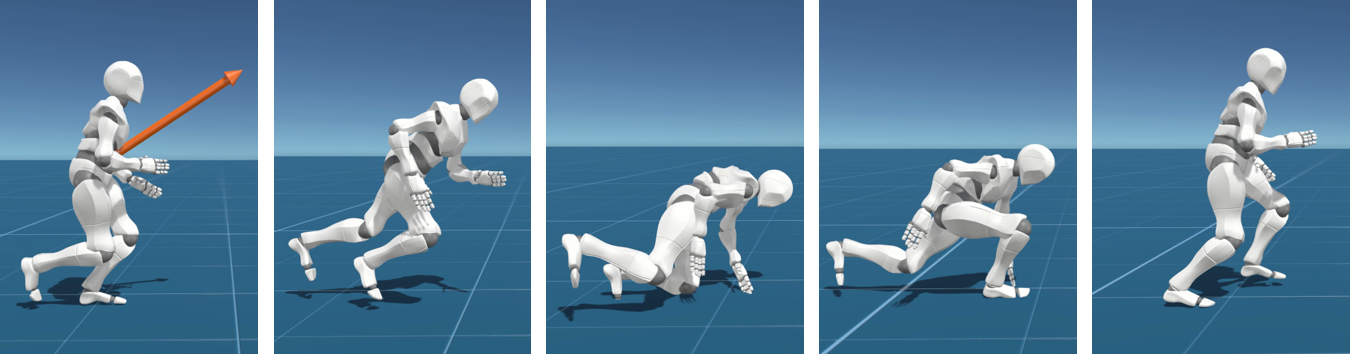}
    \caption{Push and Recovery}
    \label{Fig:recover}
  \end{subfigure} \hspace{1mm}
  \begin{subfigure}{0.48\linewidth}
    \includegraphics[width=\linewidth]{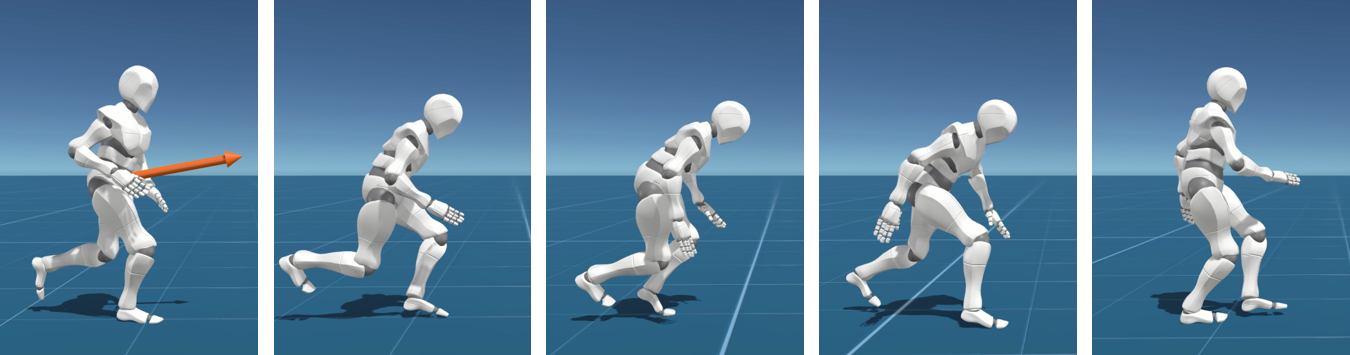}
    \caption{Keep Balance}
    \label{Fig:resist}
  \end{subfigure}
  
  \caption{Simulated character performing tasks with a high-level control policy.}
  \label{fig:interaction}
\end{figure*}

\paragraph{Heading control} 
In this task, our character is able to move and respond to user input such as changing the moving direction and speed. It automatically transits between different skills to achieve the target direction and speed (see \fig ~\ref{Fig:speed}).
It is also capable of resisting external perturbation or recovering from tumbling when moving (see \fig ~\ref{Fig:recover} and ~\ref{Fig:resist}). \fig \ref{fig:target_speed} shows the response of the character to a speed change. We can see that our character reacts quickly to adapt to the speed change and stops at last as the final target speed is zero.

\begin{figure}[t]
  \centering
  \includegraphics[width=0.7\linewidth]{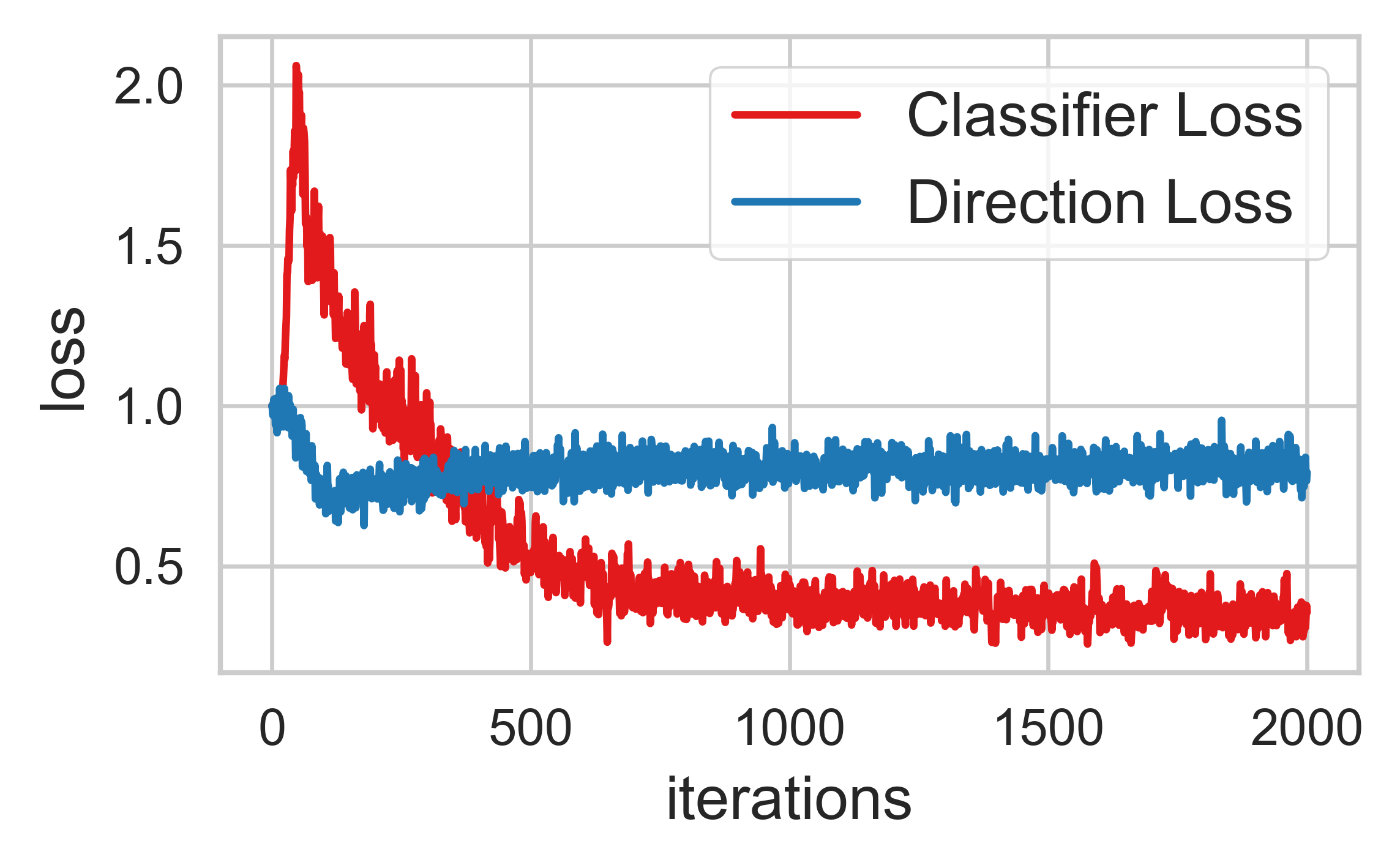}
  \caption{Typical learning curves of the skill control task. Red: the classifier loss. Blue: the direction loss. A lower classifier loss suggests that the learned policy can generate motions of the target skill. A lower direction loss indicates that the policy can accomplish the heading control task. The curves are normalized based on the values of the first iteration.}
  \label{Fig:style_loss}
\end{figure}

\fig ~\ref{Fig:target} shows a derived task of going towards a target location. We achieve this by computing the target direction and speed according to the relative distance between the character and the goal. The character can move towards the goal and stop when reaching it.

\paragraph{Steering control} 
In the steering control task, our character learns to move to the target direction while facing towards another direction. As shown in \fig\ref{Fig:walk_aside}, the character learns to walk sideways under user control. It demonstrates a more complex foot pattern compared with that used in simply moving forward.

\paragraph{Skill control}
In the skill control task, our character learns to accomplish the heading control task using a user-specified skill. Both the discriminator $D$ and classifier $C$ are modeled as neural networks with two 256-unit hidden layers. They are updated with the RAdam optimizer with the learning rates $0.0001$ and $0.01$, respectively. 
As shown in \fig\ref{Fig:Jump_turn} and \fig\ref{Fig:hop_turn}, the character can turn while maintaining the given style. 
\fig\ref{Fig:style_loss} shows the learning curve of this task. We can see that as the training processes, the policy learns the heading control first and gradually grasps the styles. 

\subsection{Comparison}
In this section, we conduct several experiments to justify our design choices and the performance of the \cvae{}.

\paragraph{Comparison with \svae{}}
One of the key components of \cvae{} is the state-conditional prior distribution. To show the effectiveness of this design, we train a different \cvae{} with the standard normal distribution $\mathcal{N}(\vect{0},\vect{I})$ as the prior distribution. This can be achieved equivalently by defining the posterior as $q(\latent_t|\stt_t, \stt_{t+1}) \sim \mathcal{N}(\vect{\mu}_q,\vect{I})$ and sampling $\latent\sim\mathcal{N}(\vect{0},\vect{I})$ when needed in the training. We refer to the \cvae{} with this configuration as the \emph{\svae{}} (VAE-NC). 

\begin{figure}[t]
  \centering  
  \begin{subfigure}{0.45\linewidth}
    \includegraphics[width=\linewidth]{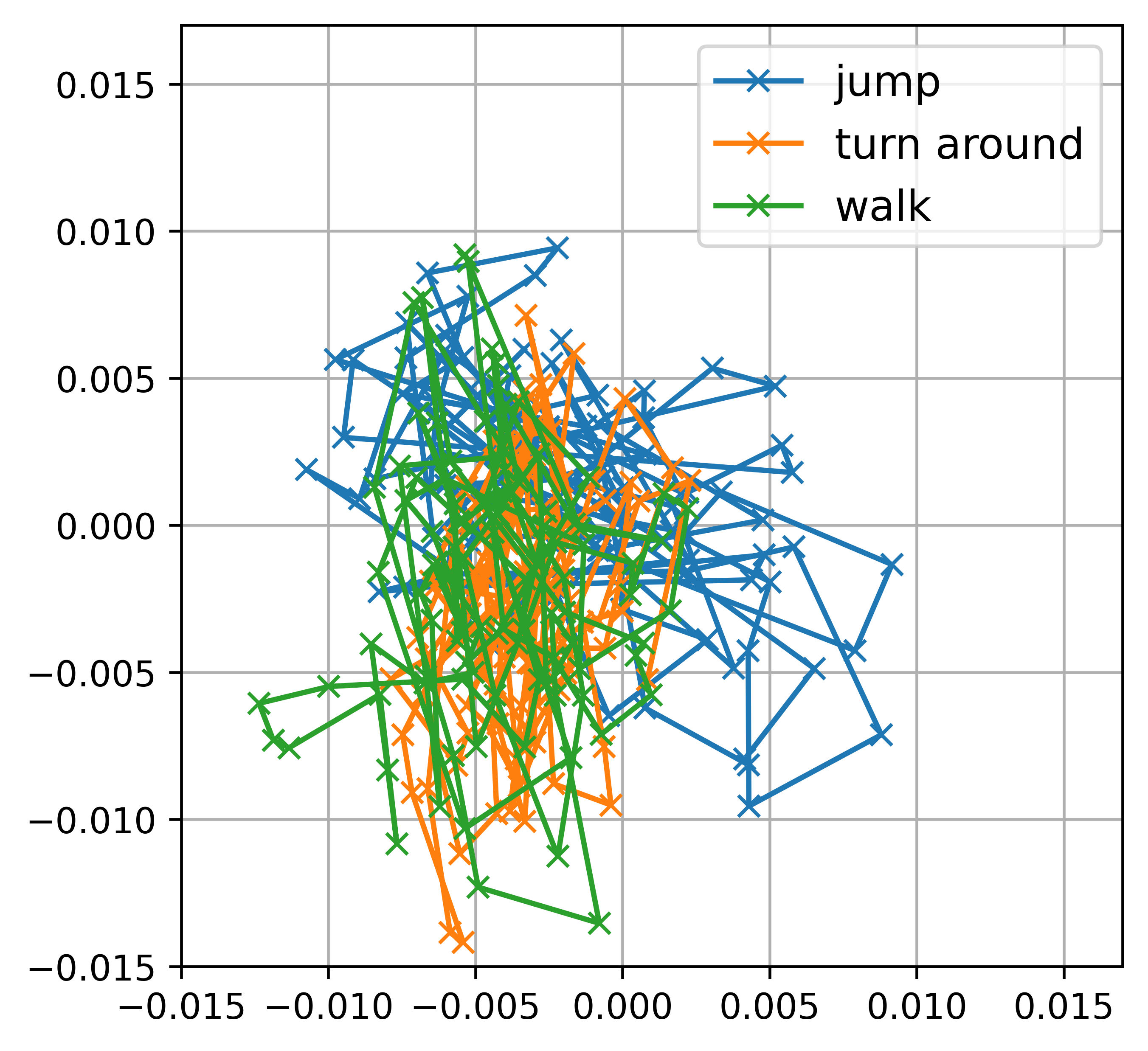}
  \caption{\svae{}}
  \end{subfigure}
  \begin{subfigure}{0.45\linewidth}
    \includegraphics[width=\linewidth]{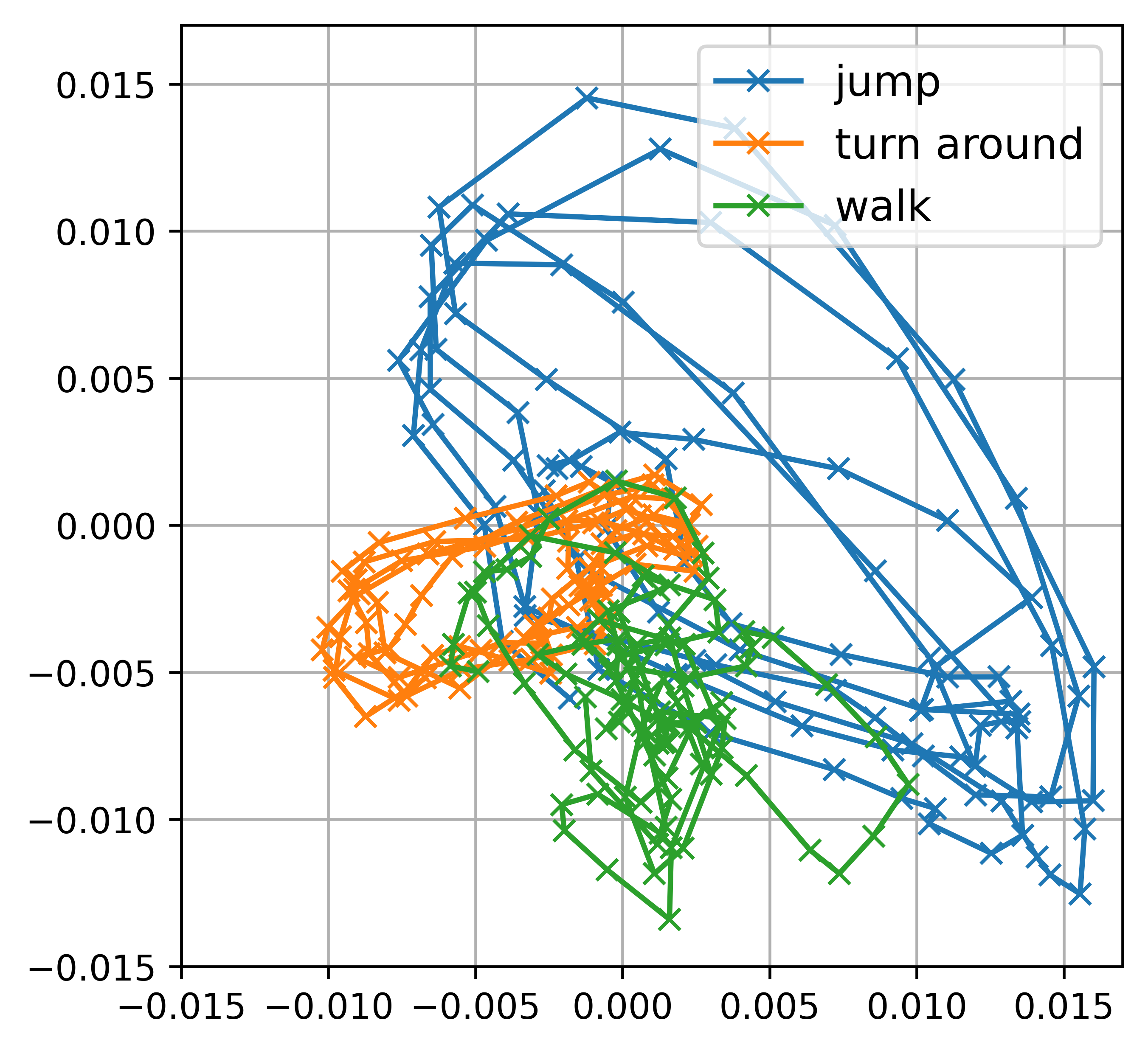}
  \caption{\cvae{}}
  \end{subfigure}
\Description{}
\caption{Visualization of different skills in the latent space.}
\label{fig:comparison_VAE_latent}
\end{figure}

\begin{figure}[t]
  \centering
  \includegraphics[width=0.7\linewidth]{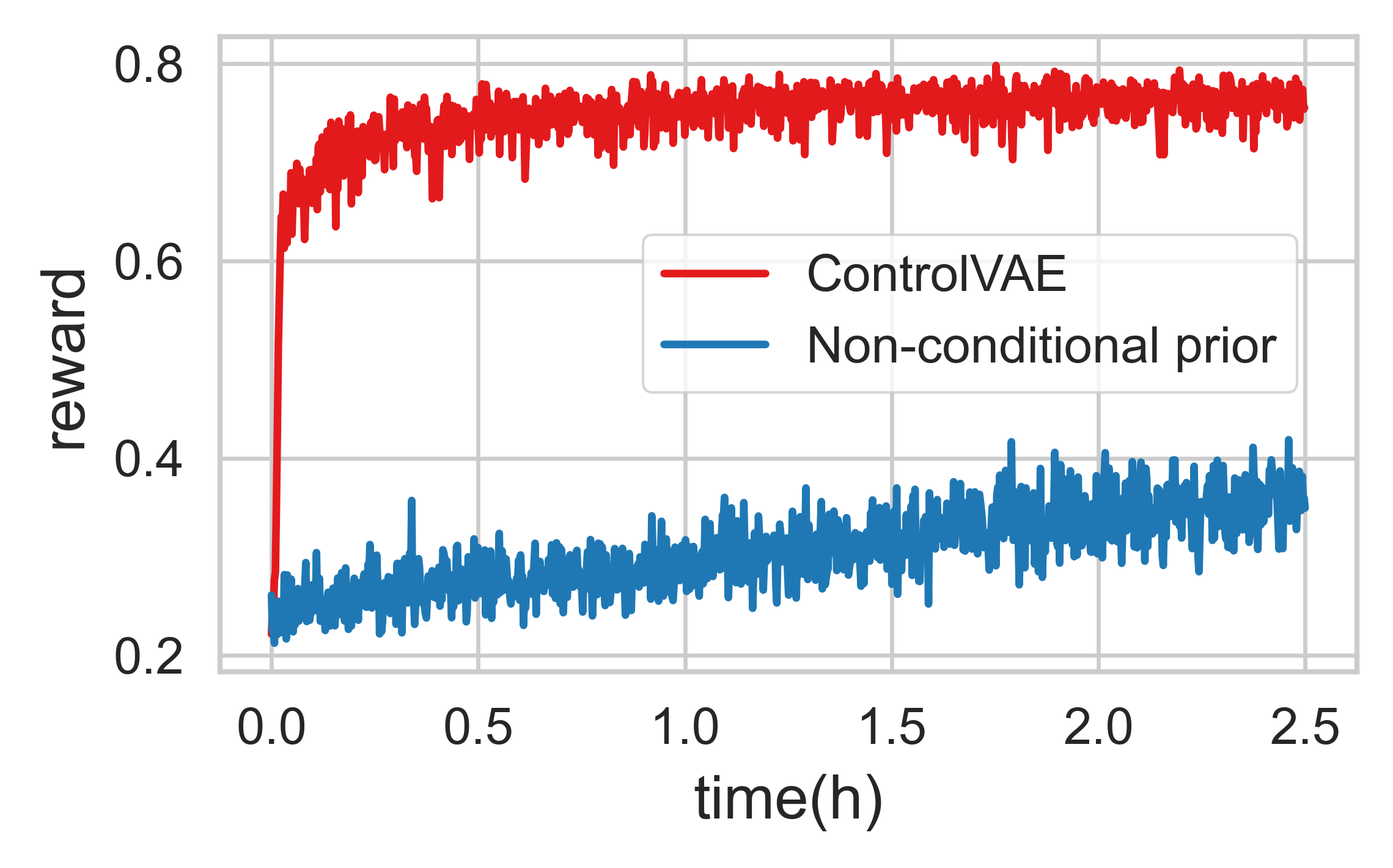}
  \Description{}
  \caption{Training curves of the heading control policies using \cvae{} and \svae{}.}
  \label{fig:comparison_VAE}
\end{figure}

\begin{figure*}[t]   
  \begin{subfigure}{0.48\linewidth}
    \includegraphics[width=\linewidth]{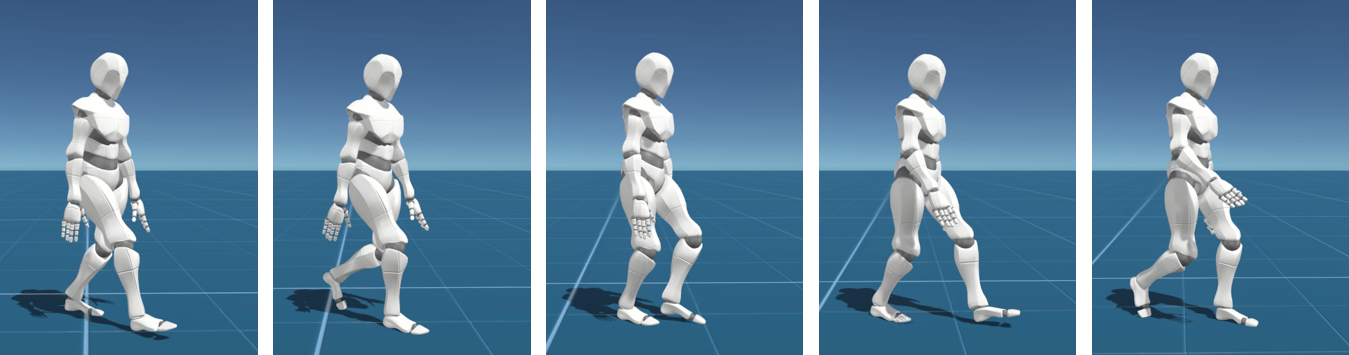}
    \caption{Walk}
  \end{subfigure}
  \hspace{1mm}
  \begin{subfigure}{0.48\linewidth}
    \includegraphics[width=\linewidth]{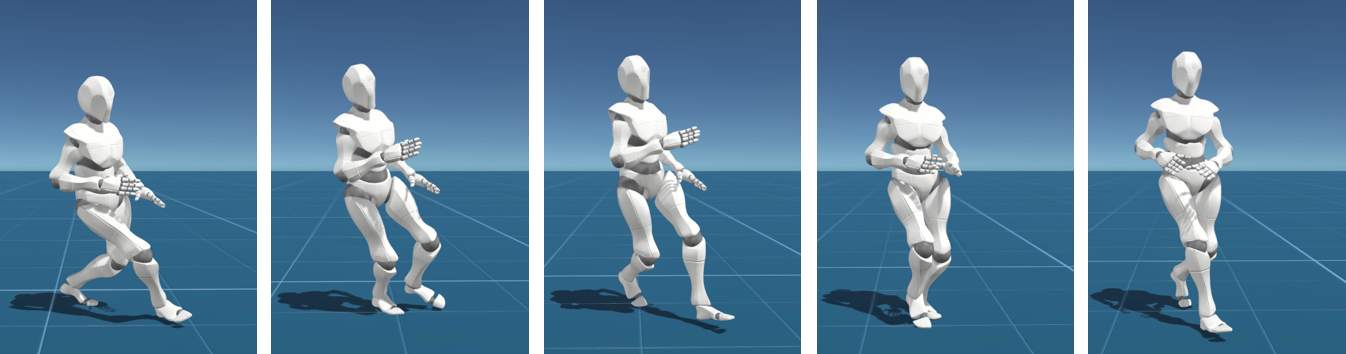}
    \caption{Run}
  \end{subfigure}

  \begin{subfigure}{0.48\linewidth}
    \includegraphics[width=\linewidth]{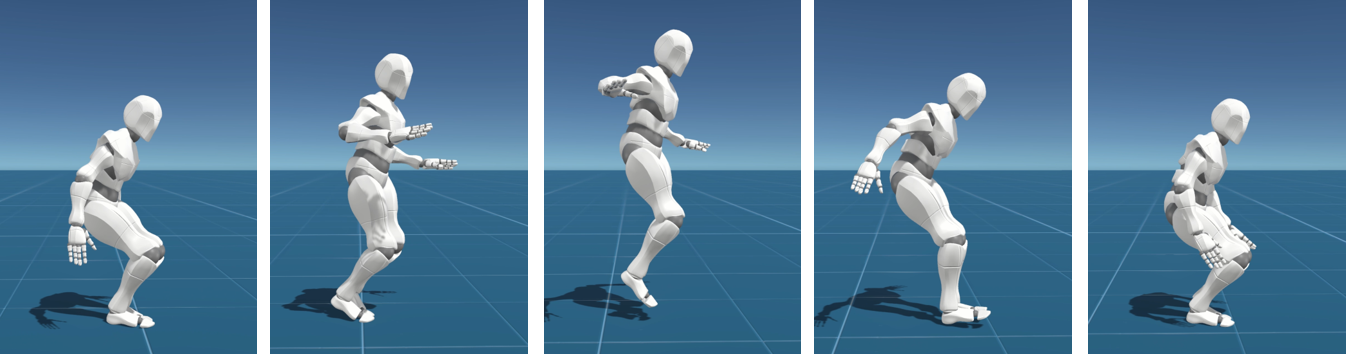}
    \caption{Jump}
  \end{subfigure}
  \hspace{1mm}
  \begin{subfigure}{0.48\linewidth}
    \includegraphics[width=\linewidth]{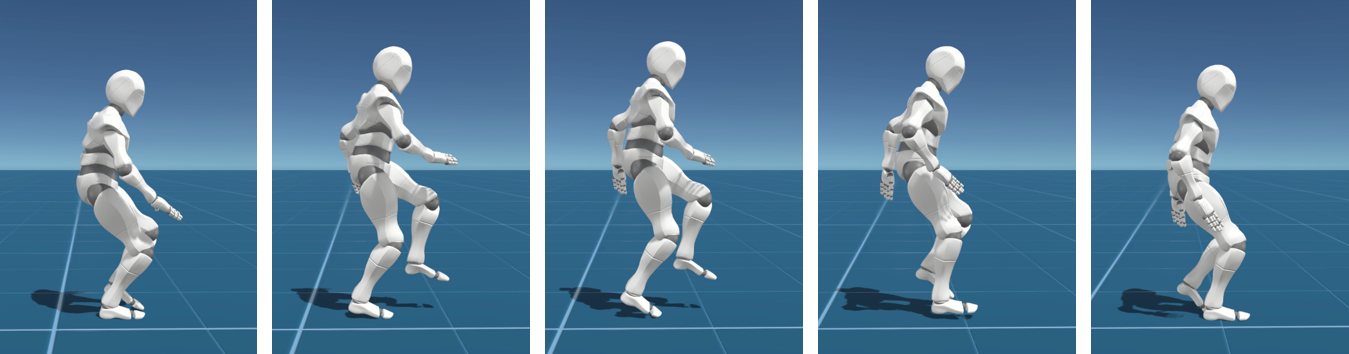}
    \caption{Hop}
  \end{subfigure}
  
  \begin{subfigure}{0.48\linewidth}
    \includegraphics[width=\linewidth]{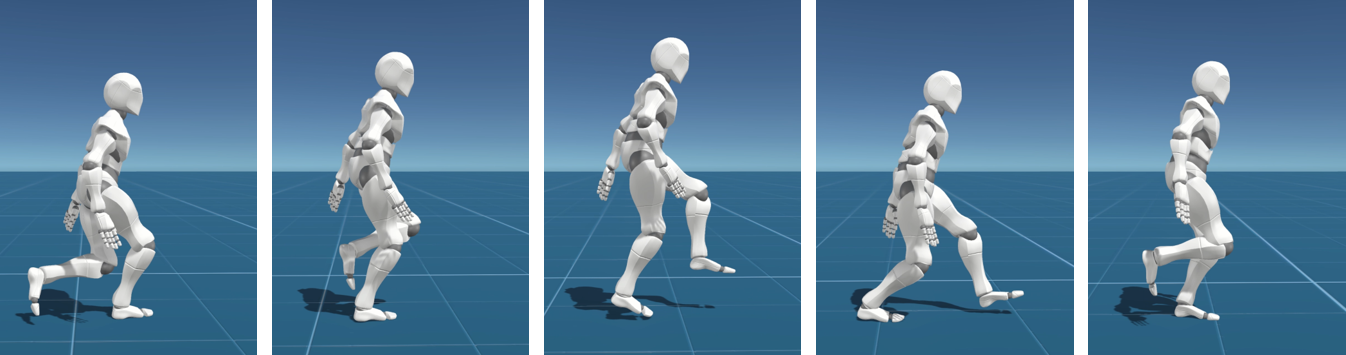}
    \caption{Skip}
  \end{subfigure}
  \hspace{1mm}
  \begin{subfigure}{0.48\linewidth}
    \includegraphics[width=\linewidth]{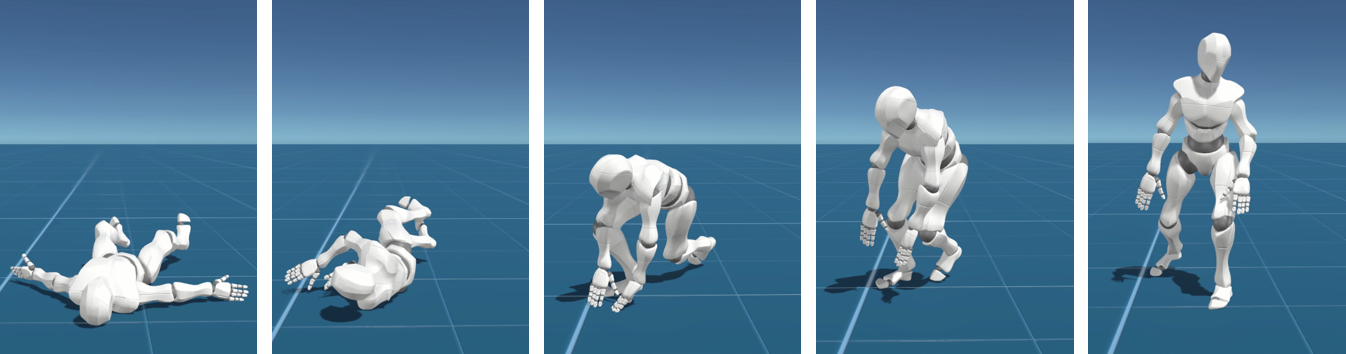}
    \caption{Getup}
  \end{subfigure}
  
  \Description{}
  \caption{Simulated character performing different skills in the skill control task}
  \label{fig:simulated_skills}
\end{figure*}

\begin{figure}[t]
  
  \centering
  \begin{subfigure}{0.48\linewidth}
    \includegraphics[width=\linewidth]{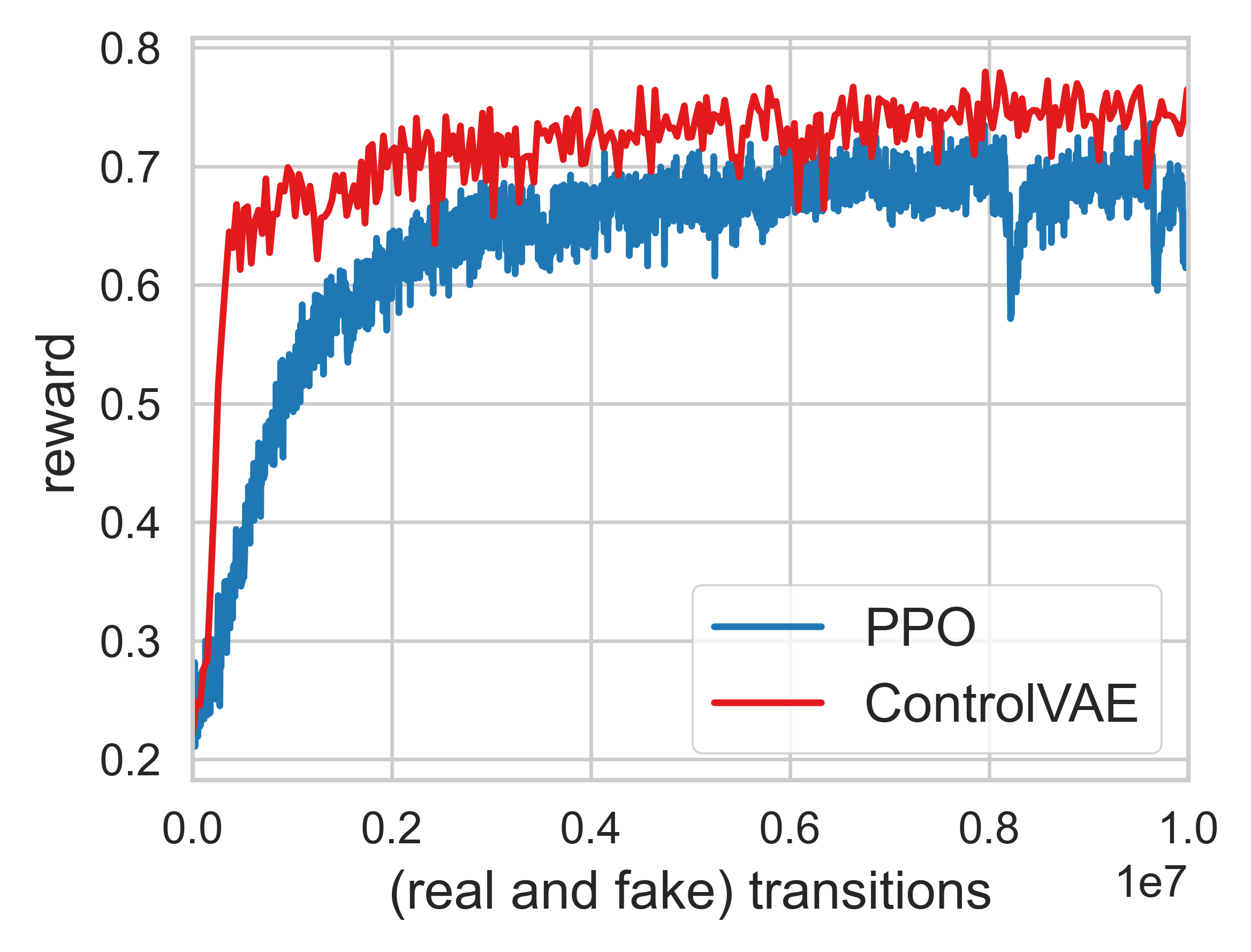}
  \end{subfigure}
  \begin{subfigure}{0.48\linewidth}
    \includegraphics[width=\linewidth]{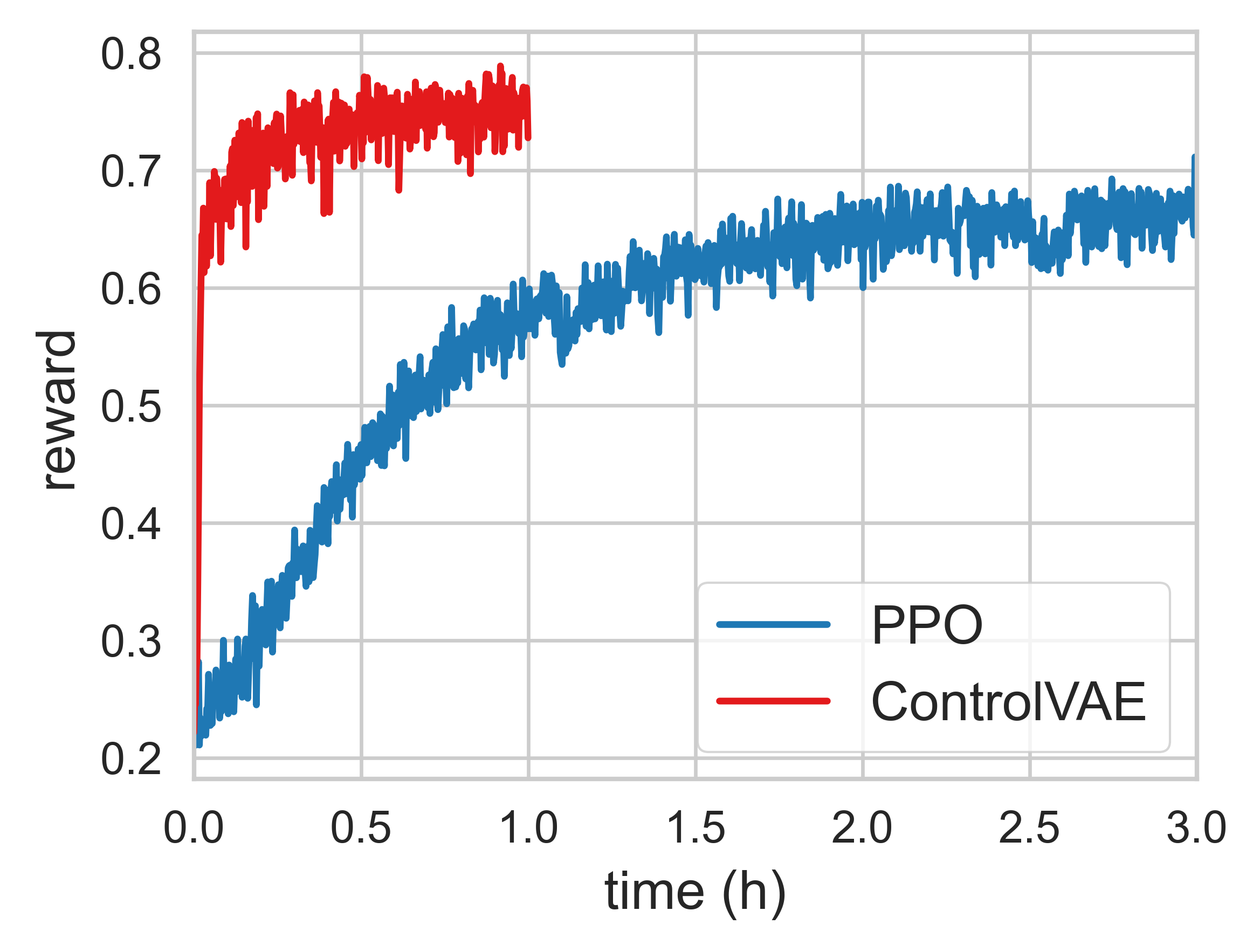}
  \end{subfigure}
  \Description{}
  \caption{Training curves of the heading control policies using model-based learning and reinforcement learning. Left: training progresses with respect to the number of transitions. Right: training progresses with respect to time.}
  \label{fig:comparison_PPO_time}
\end{figure}

\fig\ref{fig:comparison_VAE_latent} shows a comparison between the learned latent spaces of \cvae{} and \svaeShort{}, where we encode three motion clips of different skills into the latent spaces using the learned approximate posterior distributions. Generally speaking, the state-conditional prior allows each skill to be embedded into the latent space more continuously than the standard normal distribution, and the latent codes of different skills are more separated. 
This can be further justified by downstream tasks. In the {random sampling} test, the character controlled by \svaeShort{} behaves more unstably, often falls on the ground, and struggles to get up. Also in the {heading control} task, the character controlled by \svaeShort{} can hardly follow the target direction but keep turning and falling. The corresponding learning curve in \fig\ref{fig:comparison_VAE} also shows that the \svae{} performs worse than \cvae{}.


\paragraph{Comparison with reinforcement learning}

Many recent works achi-\\eve
successful hierarchical control policies using model-free reinforcement learning algorithms, where a task policy is trained on top of a learned skill latent space \cite{lingCharacterControllersUsing2020a,merelCatchCarryReusable2020,luoCARLControllableAgent2020}. Our \cvae{} also support training task-specific policies in this way. As a comparison with our model-based learning paradigm, we train the same heading control task using the PPO algorithm~\cite{schulmanProximalPolicyOptimization2017a}, which is often implemented as a model-free reinforcement learning approach and is widely used in physics-based character animation.

More specifically, the reinforcement learning algorithm optimizes a control policy $\pi$ by maximizing the expected return over all possible simulation trajectories induced by $\pi$.
To train a task-specific control policy $\policy(\latent|\stt,\task)$, we generate these trajectories by sampling skill variables $\latent\sim\policy(\latent|\stt,\task)$, converting $\latent$ into an action $\act$ using the skill-conditioned policy $\policy(\act|\stt,\latent)$ learned by \cvae{}, and then executing $\act$ in the simulation. The reward function is simply defined
$r_t = e^{-\mathcal{L}_{\task}}$,
where $\mathcal{L}_{\task}$ is the task-specific loss function of the heading control task in \eqn\eqref{eqn:heading_loss}. Note that we do not include the falling penalty $\mathcal{L}_{\eqword{fall}}$ in the reward, but instead early-terminate the trajectory when falls are detected. The control policy $\policy(\latent|\stt,\task)$ is modeled the same as the one we used in the model-based learning. 

We train the policy $\policy(\latent|\stt,\task)$ using our own implementation of the PPO algorithm, which can achieve comparable performance on simple tracking tasks with DeepMimic~\cite{pengDeepMimicExampleguidedDeep2018a}. \fig\ref{fig:comparison_PPO_time} shows the learning curve of this training, as well as the learning curve of the model-based learning on the same task. 
In general, the policy trained using model-based learning can achieve good performance quickly, while PPO implementation takes more time and transitions to reach a comparable level of performance. It should be noted that in \fig \ref{fig:comparison_PPO_time}, both the real and synthetic transitions are counted. The model-based learning process mainly consumes the synthetic transitions, which can be computed much more efficiently than the simulation due to the approximation nature of the world model and the help of modern GPU hardware. 
Our model-based learning approach can also achieve better performance with the same numbers of transitions due to the advantage of direct gradient delivery. In practice, the model-based learning can finish training with good results in half an hour, while the reinforcement learning will take a much longer time to obtain a quality control policy.

\subsection{Robustness, Scalability, and Generalization}

\begin{figure}[t]
  \begin{subfigure}{0.49\linewidth}
    \includegraphics[width=\linewidth]{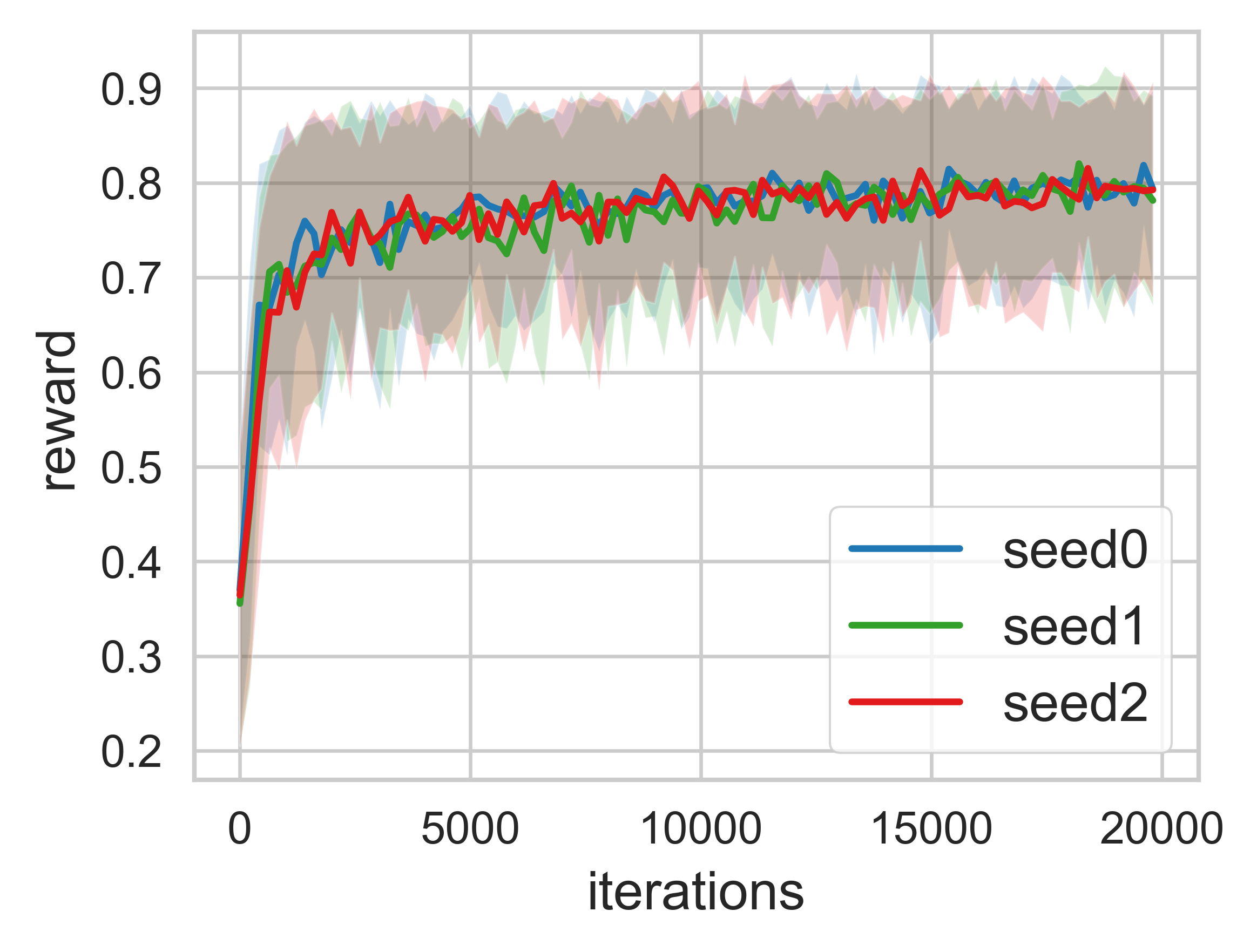}
  \Description{}
  \caption{Different random seeds. }
  \label{fig:seed_rwd}
  \end{subfigure}
  \begin{subfigure}{0.49\linewidth}
    \includegraphics[width=\linewidth]{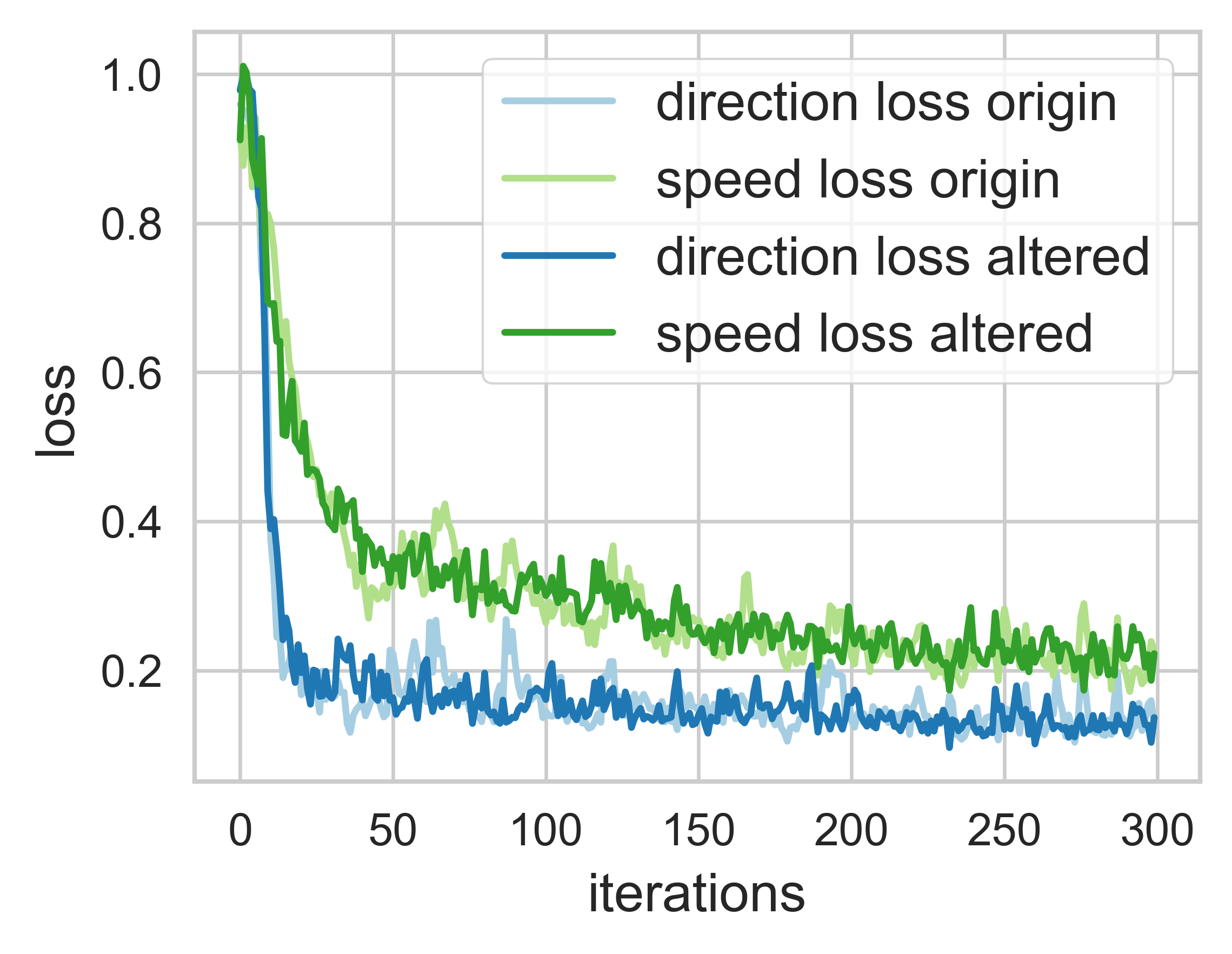}
  \Description{}
  \caption{Altered world model.  }
  \label{fig:comparison_altered_wm}
  \end{subfigure}
 
  \begin{subfigure}{0.49\linewidth}
  \includegraphics[width=\linewidth]{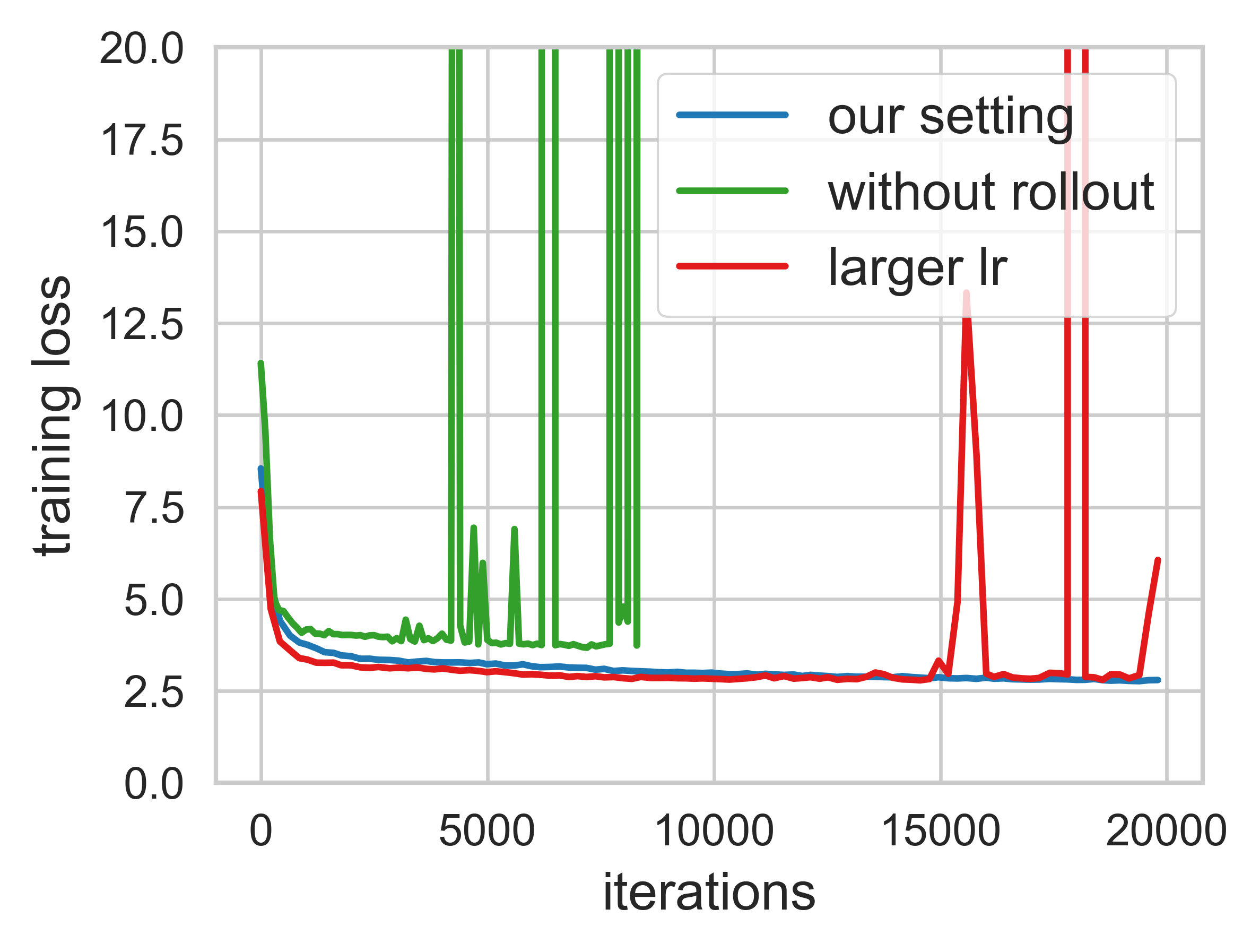}
  \Description{}
  \caption{ Different settings. }
  \label{fig:different_setting}
  \end{subfigure}
  \begin{subfigure}{0.49\linewidth}
  \includegraphics[width=\linewidth]{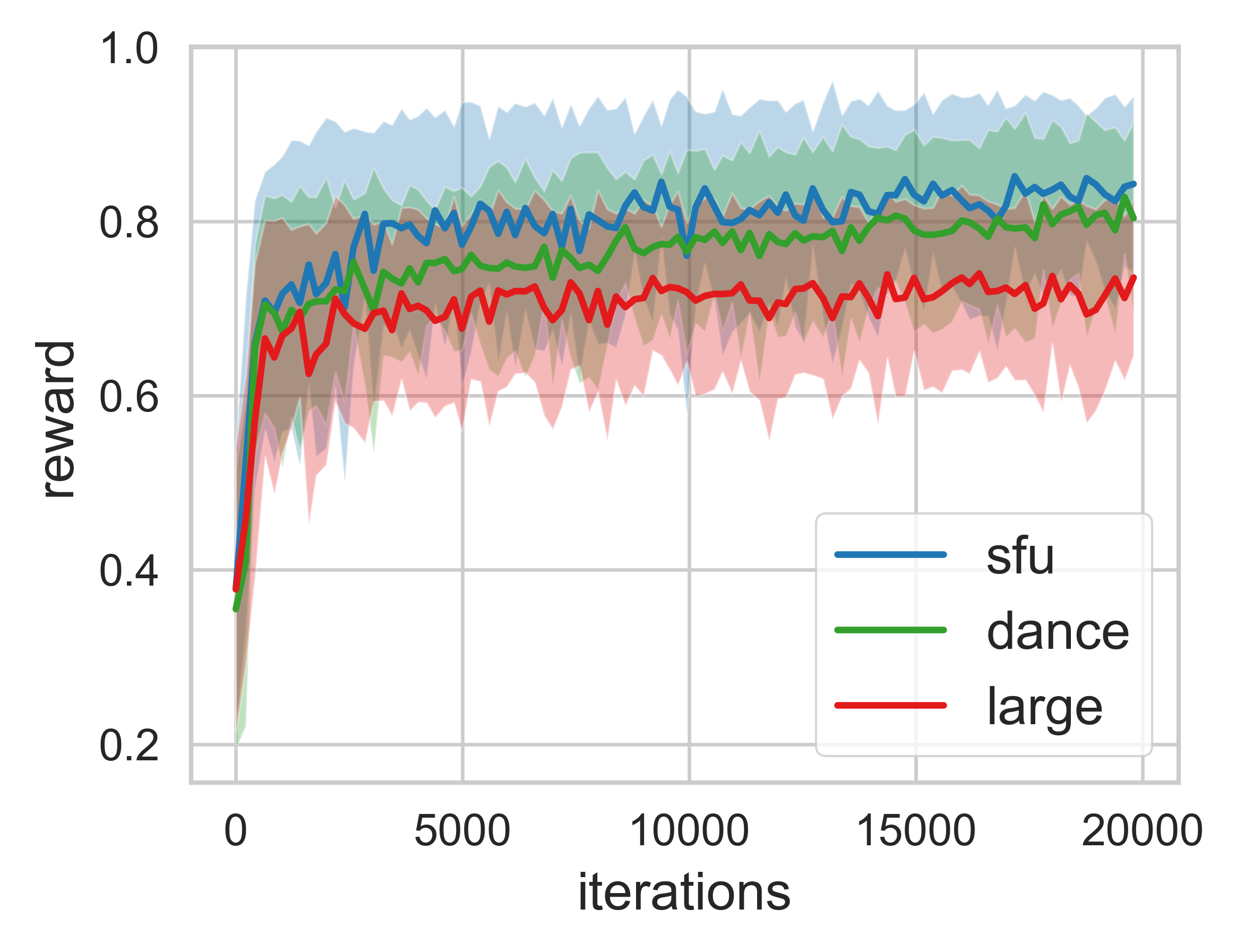}
  \Description{}
  \caption{ Different datasets. }
  \label{fig:different_datasets}
  \end{subfigure}
  \caption{\subref{fig:seed_rwd} Training \cvae{} with different random seeds. \subref{fig:comparison_altered_wm} Training heading control with the original/altered world model. We train \cvae{} and the original world model with seed $0$. The altered world model is trained with seed $1$. \subref{fig:different_setting} Training curves with different settings: using a larger learning rate and training the world model without synthetic rollouts. \subref{fig:different_datasets} Training \cvae{} on different datasets. }
\end{figure}

\begin{figure}[t]
  \centering
  \includegraphics[width=0.85\linewidth]{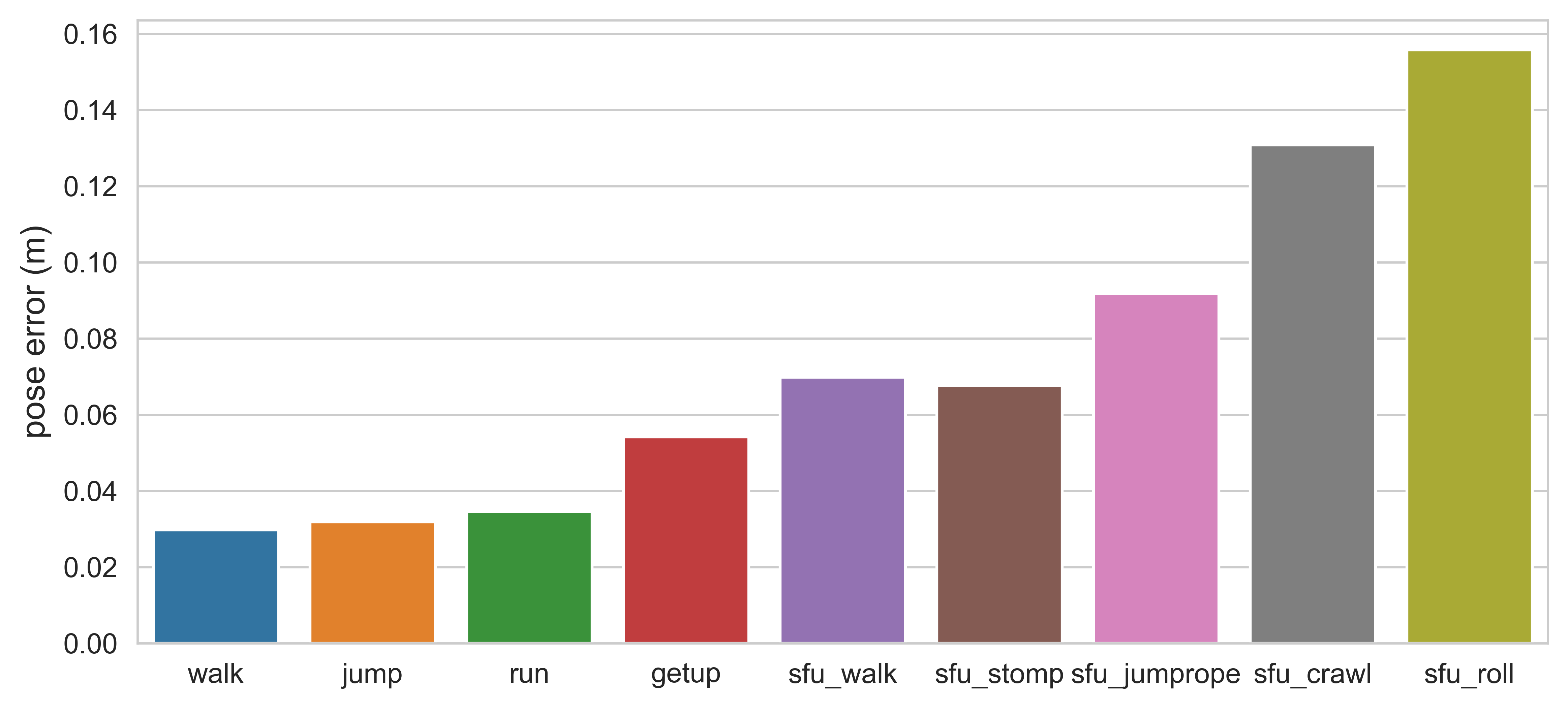}
  \Description{}
  \caption{ Tracking error of different motions. The motion with the prefix \emph{sfu} are selected form the SFU motion dataset \shortcite{SFU}, which are not used in training the \cvae{}. }
  \label{fig:tracking_error}
\end{figure}

\paragraph{Random seed} In \fig \ref{fig:seed_rwd} we show the results of training three \cvae{}s using different random seeds. It can be seen that our method performs consistently in this test. Another interesting fact is that the world models trained with different random seeds can replace each other in training the downstream locomotion tasks. To show this, we train two heading control policies using the same \cvae{} skill policy but with different world models trained with different random seeds. As shown in \fig \ref{fig:comparison_altered_wm}, the performances of the results are very close.

\paragraph{Training settings} \fig \ref{fig:different_setting} shows the effect of different training settings. We first test to train the world model without the synthetic rollouts, \ie{} $T_w = 1$. The result shows that the training process is fairly unstable and blows up early. 
In the second test, we train \cvae{} with a larger learning rate ($2e-5$). The training process converges faster than our default settings but can become unstable as the training progresses. However, even though the training loss may occasionally increase quickly, we can early-terminate the learning during the stable region, and the learned \cvae{} can still accomplish the downstream tasks.


\paragraph{Coverage} 
To test how well the learned skill embeddings cover the training skills, we let \cvae{} track all the motion sequences in the training dataset and check the accuracy of the reconstruction. We divide the long motion sequences into 2-second clips, and calculate the average tracking error between the relative position of each body to the root and that in the reference. As shown in \fig \ref{fig:tracking_error}, the learned \cvae{} can reconstruct the training motions accurately, which indicates that all the training skills are embedded in the latent space. 
We further test the ability of the learned \cvae{} in reconstructing unseen motions. Five motion clips from the SFU dataset \shortcite{SFU} are selected in this test, as shown in \fig\ref{fig:tracking_error}, which are not used in the training. As indicated in \fig\ref{fig:tracking_error}, \cvae{} can track the unseen motions that are similar to the existing skills in the training dataset accurately, such as stylized walking and stomping, albeit with some motion details missing. For the motions that are significantly different, such as rolling, the character only struggles on the ground, causing a large tracking error.



\paragraph{Dataset} We test our \cvae{} on two other datasets. One is a selected subset of the SFU mocap dataset \shortcite{SFU}, containing approximately $7$-minute diverse locomotion clips. The other is a 4-minute dance clip selected from the LaFAN dataset \cite{harvey2020robustLafanYHY}. These tests have similar training processes, and random sampling in the latent spaces creates diverse locomotion behaviors and dances. 

We further conduct an experiment on a large-scale dataset composed of 3.3 hours (after the mirror augmentation) of motions selected from the LaFAN~\cite{harvey2020robustLafanYHY} dataset, which covers most of the motion categories except for those interact with external objects or uneven terrains.
The result in \fig \ref{fig:different_datasets} shows that the training process finishes in roughly the same time as that on the small datasets but converges to a lower reward. The learned skill embeddings can still recover an input motion with a larger visual discrepancy, but the performance on the downstream tasks are significance degenerated. 



%% file: Sections/8_Discussion.tex
\section{Discussion}


In this paper, we present \cvae{}, a model-based framework for learning generative motion control policies for physics-based characters. We show that state-conditional VAEs can be efficiently trained using a model-based method, resulting in a rich and flexible latent space that captures a diverse set of skills and a skill-conditioned control policy that effectively converts each sample in the latent space into realistic behaviors in physics-based simulation.
We show that taking advantage of these generative control policies, we can generate various motion skills simply by sampling from the latent space. Task-specific control policies can be further trained to operate in this latent space, allowing a variety of high-level tasks to be accomplished using realistic motions. 

The key observation of this work is that by learning a differentiable world model, we can effectively bridge the gap between the learning of control policies and the losses defined on the simulated motion. Furthermore, the learned world model provides an effective differentiable approximation of the real simulation, which allows high-level policy to be trained efficiently using model-based learning.
We believe that our results open the door to many interesting topics, such as model-based learning of controllers using other generative models like GAN and normalizing flows.
In addition, combining model-based learning with other successful motion synthesis algorithms could be a potential way to realized physics-based motion synthesis~\cite{StarkeNeuralAnimationLayeringforSynthesizing2021}, style transfer~\cite{Aberman2020_Unpaired}, and motion in-painting~\cite{harvey2020robustLafanYHY}.

For the future work, we wish to explore the possibility of adapting the learned world model to environmental changes, such as dealing with additional objects and characters, which potentially allows us to extend a learned skill space to unseen environments. Currently, our model-based learning schemes can not handle such tasks, for which the model-free reinforcement learning methods are still needed. In addition, similar to other data-driven method, the performance of our system is still bounded by the training dataset. For example, we have encountered problems in learning a high-level policy to control the character to travel while facing towards certain directions because the corresponding motions are sparse in the dataset. 
Our \cvae{} also performs less efficiently in learning large-scale motion dataset with hours of motions, potentially due to the limited capacity of the network architecture.
Learning these sparse skills more effectively, as well as scaling up to large-scale datasets, will be a very interesting direction for future exploration.

%% file: Sections/x_ack.tex
\begin{acks}
    We thank the anonymous reviewers for their constructive comments.
    This work was supported in part by NSFC Projects of International Cooperation and Exchanges (62161146002).
\end{acks}

%% file: Sections/x_appendix.tex
\appendix

